\documentclass[fleqn,usenatbib]{mnras}
\usepackage{newtxtext,newtxmath}
\usepackage[T1]{fontenc}
\usepackage{ae,aecompl}
\usepackage{graphicx}	
\usepackage{amsmath}
\usepackage{multicol}        
\usepackage{bm}	
\usepackage{pdflscape}
\usepackage[usenames]{color}
\usepackage{soul}
\usepackage{wrapfig}
\usepackage[utf8]{inputenc}
\usepackage[section]{placeins}
\usepackage{float}
\usepackage{pgfplots}
\usepackage{natbib}
\usepackage[nottoc]{tocbibind}
\usepackage{color,soul}
\usepackage{enumerate}
\usepackage{caption}
\usepackage{subcaption}
\usepackage{verbatim}

\newcommand{\jwm}[1]{{\color{red} #1}}

\title{A Force Explosion Condition for Spherically Symmetric Core-collapse Supernovae}
\author[M. Gogilashvili \& J.W. Murphy]{
Mariam Gogilashvili,$^{1}$ \thanks{Email: mg18u@my.fsu.edu}
Jeremiah W. Murphy,$^{1}$ \thanks{Email: jwmurphy@fsu.edu}
\\
$^{1}$Department of Physics, Florida State University, 77 Chieftan Way, Tallahassee, FL 32306, USA}

\begin{document}
\label{firstpage}
\pagerange{\pageref{firstpage}--\pageref{lastpage}}
\maketitle

\begin{abstract}
  Understanding which stars explode leaving behind neutron stars and which stars collapse forming black holes remains a fundamental astrophysical problem.  We derive an analytic explosion condition for spherically symmetric core-collapse supernovae.  The derivation starts with the exact governing equations, considers the balance of integral forces, includes the important dimensionless parameters, and includes an explicit set of self-consistent approximations. The force explosion condition is  $\tilde{L}_\nu\tau_g - 0.06 \tilde{\kappa} > 0.38$, and only depends upon two dimensionless parameters.  The first compares the neutrino power deposited in the gain region with the accretion power, $\tilde{L}_\nu \tau_g = L_{\nu} \tau_g R_{\rm NS}/ ( G \dot{M} M_{\rm NS})$.  The second, $\tilde{\kappa} = \kappa \dot{M} / \sqrt{G M_{\rm NS} R_{\rm NS}}$, parameterizes the neutrino optical depth in the accreted matter near the neutron-star surface. Over the years, many have proposed approximate explosion conditions: the critical neutrino-luminosity, ante-sonic, and timescale conditions.  We are able to derive these other conditions from the force explosion condition, which unifies them all.  Using numerical, steady-state and fully hydrodynamic solutions, we test the explosion condition. The success of these tests is promising in two ways.  One, the force explosion condition helps to illuminate the underlying physics of explosions.  Two, this condition may be a useful explosion diagnostic for more realistic, three-dimensional radiation hydrodynamic core-collapse simulations. 
\end{abstract}

\begingroup
\let\clearpage\relax
\endgroup
\newpage

\section{Introduction}

While nearly all massive stars end their lives with core collapse, sometimes this collapse initiates an explosion, core-collapse supernovae (CCSNe)  \citep{LI2011,HORIUCHI2011}, and sometimes they fail to explode, forming black holes \citep{FISCHER2009, OCONNOR2011}.  CCSNe are one of the most energetic explosions; the kinetic energy released during the explosion plays an important role in star formation in the ISM \citep{MACLOW2004}.
Understanding which stars explode and which collapse to a black hole is important in understanding the final fates of massive stars. Being able to predict which stars explodes directly impacts our ability to predict neutron star and black hole distributions and ultimately impacts our understanding of compact merger detections via gravitational waves.  Furthermore, knowing the distribution of stars that explode is important in being able to predict nucleosynthesis \citep{WOOSLEY2002, DIEHL2021}.

Before the iron (Fe) core collapses, it is mostly supported by electron degeneracy pressure, and once the Fe core reaches the Chandrasekhar Mass, the Fe core collapses on a dynamical timescale ($\sim$0.1 s).  This collapse is aided by electron capture and photo-dissociation of the Fe nuclei.  The core bounces at nuclear densities due to the residual strong force, forming a proto-neutron star.  This bounce  launches a shock wave, which soon stalls due to energy losses from photo-dissociation of the heavy nuclei and neutrino emission. \citep{HILLEBRANDT1981,MAZUREK1982,MAZUREK82}. If the shock remains stalled, the collapsing star continues to accrete onto the proto-neutron star, leading to a black hole and a failed supernova  \citep{FISCHER2009,OCONNOR2011}.

The primary challenge in CCSN theory is understanding how and under what conditions the stalled shock revives into an explosion.
 \citet{COLGATE1966} and \citet{BETHE1985} showed that neutrinos can transport thermal energy from the proto-neutron star into the gain region, heat the material behind the shock.  In some circumstances this may lead to shock revival. This delayed-neutrino mechanism fails in most of the one-dimensional (1D) simulations; of these  spherically symmetric simulations, only the lower mass progenitors explode \citep{LIEBENDORFER2001A,LIEBENDORFER2001b,LIEBENDORFER2005,
RAMPP2002,BURAS2003,BURAS2006,THOMPSON2003,KITAURA2006,MULLER2017,
RADICE2017}. While 1D simulations fail, multi-dimensional simulations tend to explode \citep{BENZ1994,HERANT1994,BURROWS1995,JANKA1995, LENTZ2015, BRUENN2006, BRUENN2016,MULLER2015, MULLER2019, VARTANYAN2018, SKINNER2019, STEINER2013, VARTANYAN2019, VARTANYAN2021}. Multi-dimensional simulations have and will always likely be important tools in understanding how and which stars explode. However, the 3D simulations are computationally expensive ($\sim$10 million cpu-hours or $\sim$\$1,000,000 in electricity costs per run), making it difficult to run large parameter studies.  These non-linear, multi-physics simulations also do not readily present an intuitive understanding of the explosion mechanism. An approach that complements the simulations is to understand the underlying physics with analytic or semi-analytic calculations.  These analytic calculations can provide subgrid models to speed up the simualtions \citep{MABANTA2019}  and they can provide a deeper understanding of the explosion conditions. 

One semi-analytic explosion condition that has made a significant impact is the critical neutrino-luminosity condition.
 \citet{BURROWS1993} noticed that there are no steady-state, stalled shock solutions above a critical neutrino luminosity ($L_{\nu}$); they suggested that this curve represents an explosion condition.  Using this condition as a diagnostic tool, \citet{YAMASAKI2005} explored how rotation would reduce the critical neutrion luminosity.  \citet{MURPHY2008} showed that the neutrino critical condition is indeed consistent with explosions in spherically symmetric simulations, and they used this condition to quantify that the critical neutrino luminosity condition is $\sim$30\% lower in multi-dimensional simulations.  Subsequent simulations confirmed these results \citep{HANKE2012, DOLENCE2013, HANDY2014, FERNANDEZ2015}.   Later  \citet{MABANTA2018} included the subgrid convection model of \citet{MURPHY2013} in the semi-analytic calculations and showed that the subgrid convection model reduces the critical neutrino luminosity by $\sim$30\%.  They also showed that heating by turbulent dissipation is responsible for most of this reduction.

Over the years, several have suggested more heuristic explosion conditions.  These are not derived, rather they are suggestions based upon either informed intuition or empirical results.
For example, \citet{THOMPSON2000} proposed a time-scale explosion condition; explosions succeed if the heating timescale within the gain region is short compared to the advection time scale through the gain region, $\tau_{\rm adv}/\tau_{q}>1$. While these time-scale conditions are intuitive, they are inadequate predictors of explosion for two reasons.  One, without a formal definition, there are many ways to to define the time scales \citep{MURPHY2008}.  Two, when compared to simulations, they are only order-of-magnitude accurate. More recently, \citet{OCONNOR2011} proposed  a compactness parameter as an explosion condition.  By forcing explosions by adjusting the neutrino heating, they noticed a rough correlation between the success of explosion and the compactness of the Fe core before collapse, $\xi_M=\frac{M/M_\odot}{R(M_{\rm bary}=M)/1000km}\bigg |_{t=t_{\rm bounce}}$.  While this parameter is easy to calculate from the progenitors, CCSN simulations show that it is also an order-of-magnitude estimate \citep{ERTL2016}.

\citet{KESHET2012} attempted to derive the neutrino-luminosity critical condition of \citet{BURROWS1993}.  Ultimately deriving an analytic critical condition requires analytically modeling a boundary value problem, with the bottom boundary being the Netron Star (NS) and the upper boundary being the stalled shock. Their particular strategy focused on showing that both the neutrino optical depth, $\tau$, and the density at the surface of the NS, $\rho_\nu$, have a maximum for a specific shock radius. They found that this maximum is a monotonic function of the neutrino luminosity; above a critical neutrino luminosity, there are no stalled shock solutions. To provide a closed form for the boundary value problem, they approximated the neutrino luminosity as black-body radiation at the neutrinosphere.   This particular assumption allows for a closed analytic solution; however, it also excludes the need for using the energy conservation equation. Hence, this critical condition does not explicitly include the neutrino power deposited in the gain region, which is believed to play an important role in the explodability. As yet, neither these assumptions nor these results have been validated with simulations.  

In another attempt to understand the physics behind the neutrino-luminosity critical condition, \citet{PEJTA2012} introduced the ante-sonic condition. First, they explored isothermal accretion and showed that there is a maximum sound speed above which steady-state solutions do not exist.  Suspecting similar physics would apply to adiabatic flows, they then used numerical solutions to show that there is roughly an ante-sonic condition for adiabatic accretion as well. The ante-sonic critical condition for the adiabatic flows is $c_s^2/\varv_{\rm esc}^2 \simeq 0.19 $, where $c_s$ is the adiabatic sound speed and $\varv_{\rm esc}$ is the escape velocity at the shock. Later, \citet{RIVES2018} analytically derived an antesonic condition of \citet{PEJTA2012}.  \citet{RAIVES2021} extended the ante-sonic condition for the flows with rotation and turbulence.  Again, the ante-sonic condition provides more insight into the explosion conditions, but it has yet to demonstrate precise consistency with simulations.

 More recently, \citet{MURPHY2017} used the same semi-analytic technique as \citet{BURROWS1993} to propose an integral condition for explosion.  By integrating the equations of hydrodynamics, they derive an algebraic connection between the steady-state, post-shock structure and the shock velocity.  They find that the dimensional integral of the momentum equation, $\Psi$, plays an important role in whether stalled-shock solutions exist.  $\Psi$ is an integral that measures the balance of pressure, ram pressure, and gravity; in other words, it is an integral measure of the balance of forces.  In particular, the balance of forces parameter, $\Psi$, and the shock velocity are related by $\varv_s/\varv_{\rm acc}\approx -1+\sqrt{1+\Psi}$.  The physical parameters of the problem, $L_{\nu}$, $M_{\rm NS}$, $R_{\rm NS}$, and $\dot{M}$, determine the post-shock solutions.  If it is possible to find a post-shock solution with $\Psi = 0$, then there exits a stalled solution, $\varv_s = 0$.  Above the critical neutrino luminosity condition, only $\Psi > 0$ solutions exist.  \citet{MURPHY2017} showed that this force explosion condition is consistent with the critical neutrino luminosity condition and one-dimensional simulations.  This integral condition also shows that the critical condition depends upon more than just the neutrino luminosity or the accretion rate.  They show that the force explosion condition corresponds to a critical hypersurface where the physical dimensions are $L_{\nu}$, $M_{\rm NS}$, $R_{\rm NS}$, and $\dot{M}$.  \citet{MURPHY2017} also found the force explosion condition to be a useful explosion diagnostic for simple light-bulb simulations.  While the force explosion condition is consistent with simple simulations, and it promises to be a useful explosion diagnostic, there remains a couple of problems.  For one, the work of \citet{MURPHY2017} relied on numerical steady-state solutions; an analytic derivation would provide greater clarity on why the explosion condition exists.  Second, \citet{MURPHY2017} used numerical steady-state solutions to calculate the explosion diagnostic for the hydrodynamic simulations.  Such a procedure is too cumbersome for wide-spread use; an analytic explosion diagnostic would be easier to use among the community. 

In this manuscript, we complete the process started in  \cite{MURPHY2017} and derive the analytic force condition for explosion.
The structure of this manuscript is as follows. The derivation in section~\ref{analytic critical condition} is self-consistent in that it starts from the fundamental hydrodynamics equations and includes explicit assumptions and approximations.  This derivation also begins with identifying the important dimensionless parameters of the problem.  In doing so, we show that the number of parameters in the explosion condition reduces from four physical parameters to two dimensionless parameters.  Sections~\ref{ODE}~\&~\ref{CUFE} verify and validate the analytic force condition using two techniques. In the first, we show that the analytic explosion condition reproduces the numerical steady-state explosion conditions of \citet{BURROWS1993} and \citet{MURPHY2017}.  In the second, we show that the analytic explosion condition is an accurate explosion diagnostic for one-dimensional, light-bulb simulations.  In section~\ref{Discussion}, we use the force explosion condition to derive other more approximate explosion conditions, and we identify the missing components in these other approximate conditions.  In section~\ref{Conclusion}, we summarize our results and suggest ways that simulations can test  analytic force condition.  
\section{Analytic critical condition}\label{analytic critical condition}

\subsection{Defining First Principles}
To derive the explosion condition from first principles, we begin with conservation of mass, momentum, and energy:
\begin{equation}
\frac{\partial \rho}{\partial t}+\boldsymbol{\nabla}\cdot(\rho \varv)=0 \, ,
\label{mass}
\end{equation}
\begin{equation}
\rho\frac{\partial\varv}{\partial t}+\rho(\boldsymbol{\varv} \cdot \boldsymbol{\nabla})\boldsymbol{\varv}=-\boldsymbol{\nabla}p - \rho \boldsymbol{\nabla}\Phi \, ,
\label{momentum}
\end{equation}
\begin{equation}
\frac{\partial (\rho E)}{\partial t}+\boldsymbol{\nabla} \cdot \left(\rho \boldsymbol{\varv} (\varepsilon+\frac{p}{\rho}+\frac{\varv^2}{2}+\Phi)\right)= \rho q_\nu-\rho q_c \, ,
\label{energy cont}
\end{equation}
 where $\rho$ is the mass density, $\varv$ is the velocity, $p$ is the pressure, $\Phi=-\frac{G M_{\rm NS}}{r}$ is the simple Newtonian potential, $q_\nu=\frac{L_\nu \kappa}{4\pi r^2}$ is the local specific neutrino heating and $q_c=C_0 \left(\frac{T}{T_0}\right)^6$ is the local specific cooling \citep{JANKA2001}. 

The first assumption is that the stalled solutions satisfy stationary equations ($\partial_t=0$) \citep{BURROWS1993} and that explosions occur when the region between the NS and shock can no longer support steady solutions \citep{MURPHY2017}.  \cite{MURPHY2017} showed that the balance, or rather imbalance, of forces (ram pressure, pressure and gravity) is a guide to understanding the conditions for explosion.  In particular, the steady-state, integrated momentum equation represents the balance of forces:

\begin{equation}
\int_{\delta V} \boldsymbol{\nabla}\cdot(\rho \boldsymbol{\varv}\boldsymbol{\varv}) dV=-\int_{\delta V}\boldsymbol{\nabla}p dV+ \int_{\delta V}\rho \boldsymbol g dV \, ,
\label{int_momentum}
\end{equation}
 where $\boldsymbol{g}=-\boldsymbol{\nabla}\Phi$ and $\delta V$ is the shell covering the region from the neutron star to the shock (shock is included in the integration area). To simplify our calculations, we assume spherical symmetry; Equation~(\ref{int_momentum}) in spherical symmetry has the following form:
\begin{equation}
4\pi\int_{R_{\rm NS}}^{R_s+\epsilon}\frac{\partial}{\partial r}(\rho \varv^2 r^2) dr=-4\pi\int_{R_{\rm NS}}^{R_s+\epsilon} \frac{\partial p}{\partial r} r^2dr+ 4\pi\int_{R_{\rm NS}}^{R_s+\epsilon}\rho g r^2dr \, ,
\end{equation}
where $\epsilon\rightarrow 0+$ is an infinitesimal number which is in the bounds to include the shock in the integration. A useful parameter is the integrated balance of forces:
\begin{equation}
\label{psi1}
\Psi=-4\pi\int_{R_{\rm NS}}^{R_s+\epsilon}(\frac{\partial}{\partial r}(\rho \varv^2 r^2)+\frac{\partial p}{\partial r} r^2- \rho g r^2) dr \, .
\end{equation}    
Taking the limit $\epsilon \rightarrow 0+$ provides an integral equation which includes boundary conditions at the NS surface and at the shock.

Formally, the integral could include the entire subsonic region below the shock.  Hence, the lower boundary in eq.~(\ref{psi1}) could have been $r=0$.  However, there are two reasons that it is more convenient to set the lower boundary to $r=R_{\rm NS}$.  First, the accretion rate changes character at the neutrino sphere (or NS radius).  Above $R_{\rm NS}$, the mass accretion rate is relatively constant with radius.  At $R_{\rm NS}$, the accreted matter tends to settle and the mass accretion rate drops to zero there and below.   Second, once the stars explode, matter below $R_{\rm NS}$ remains bound, and matter above tends to become unbound.  Hence, the matter interior to $R_{\rm NS}$ sets the gravitational scale for explosion.

\citet{MURPHY2017} showed that the shock velocity is related to these balance of forces via $\varv_s/\varv_{+\epsilon} = -1 + \sqrt{1 + \overline{\Psi}}$.  $\varv_{+\epsilon}$ is the velocity of matter accreting onto the shock, and $\overline{\Psi}=\Psi/(\rho_{+\epsilon}\varv^2_{+\epsilon}r_s^2)$ is a dimensionless representation of the integrated forces $\rho_{+\epsilon}\varv^2_{+\epsilon}r_s^2$ is the ram pressure force on to the shock.  The explosion condition corresponds to conditions in which $\overline{\Psi} > 0$ always.  Therefore, our goal is to derive an analytic expression for $\overline{\Psi}$ and $\overline{\Psi} > 0$. Later in the derivation, we choose a different normalization for $\Psi$ which is a more natural normalization (see eq.~{\ref{tPsi}}).

\subsection{An Analytic Expression for ${\Psi}$}
\label{subsection: simple model}

\begin{figure}
    \centering
    \includegraphics[scale=0.5]{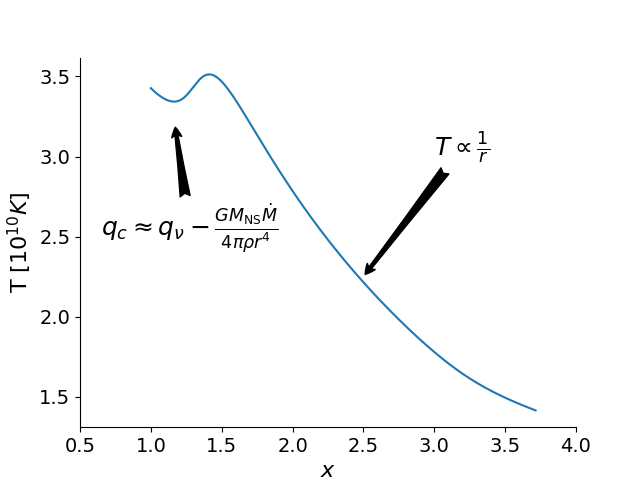}
    \caption{ Temperature as a function of radius for a stalled-shock solution to the ODEs (eqs.~(\ref{mass})-~(\ref{energy cont})).  In the gain region, the temperature roughly scales as $T \propto 1/r$.  Neutrino heating and advection cause the profile to be slightly steeper than $1/r$. In the cooling region, cooling is roughly balanced by neutrino heating and gravitational heating.  As a result, the temperature gradient is roughly zero in this region.  It is not exactly zero, but this approximation enables a relatively accurate analytic approximation.  Figure~\ref{fig_profiles} shows that the approximation $dT/dr \sim 0$ within the cooling region is also valid in one-dimensional hydrodynamic simulations.}
    \label{TvsR}
\end{figure}

To derive an analytic explosion condition from eq.~(\ref{psi1}), we propose a simple model.  One of the main purposes of this model is to propose self-consistent density and pressure profiles that enable a closed-form solution for the critical condition and hence the integral.

$\Psi$ determines whether the shock contracts, stalls, or expands (i.e. when $\Psi < 0$ then the shock contracts, $\varv_s < 0$, $\Psi = 0$ corresponds to a stalled shock, and $\Psi > 0$ corresponds to shock expansion, $\varv_s > 0$).  \citet{MURPHY2017} showed that if $\min(\Psi) > 0$, then the only possible solutions are $v_s > 0$.  Therefore, $\min(\Psi) > 0$ represents an integral condition for explosion.  To facilitate turning eq.~(\ref{psi1}) into an analytic explosion condition, we introduce the dimensionless variables and make several approximations.

The NS at the lower boundary determines the gravity and neutrino flux, and the mass accretion rate onto the shock determines the ram pressure and the density profile.  Therefore, a few important physical parameters are the NS mass, $M_{\rm NS}$, NS radius, $R_{\rm NS}$, and the mass accretion rate, $\dot{M}$.  Given these physical parameters, the dimensionless variables of the problem are:
\begin{eqnarray}
x&=&\frac{r}{R_{\rm NS}} \, , \\
\tilde{\rho}&=&\frac{\rho R_{\rm NS}^2}{\mid \dot{M} \mid}\sqrt{\frac{GM_{\rm NS}}{R_{\rm NS}}} \, , \\
\tilde{\varv}&=&\varv \sqrt{\frac{R_{\rm NS}}{G M_{\rm NS}}} \, , \\
\tilde{\varepsilon}&=&\frac{\varepsilon R_{\rm NS}}{GM_{\rm NS}} \, , \\
\tilde{p}&=&\frac{p R_{\rm NS}^2}{\mid \dot{M}\mid \sqrt{\frac{GM_{\rm NS}}{R_{\rm NS}}}} \, , \\
\tilde{\Psi}&=& \frac{\Psi}{\mid \dot{M}\mid \sqrt{\frac{GM_{\rm NS}}{R_{\rm NS}}}} \, .
\label{tPsi}
\end{eqnarray}
Given, the definition of $x$, the dimensionless shock radius is $x_s = R_s/R_{\rm NS}$.

\begin{figure}
\includegraphics[width=\columnwidth]{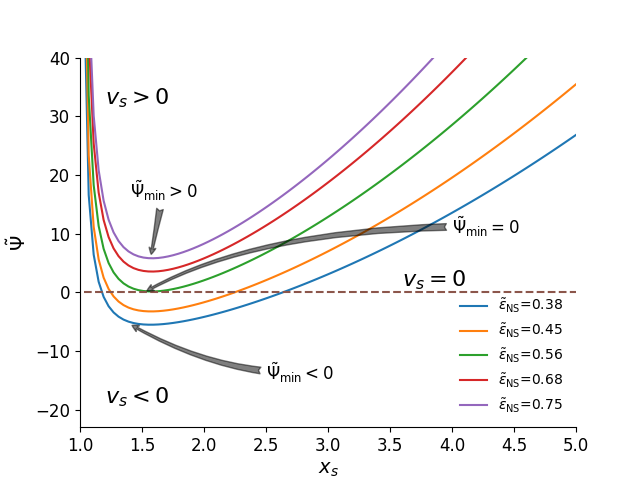}
\caption{ Analytic solutions for the dimensionless integrated force balance, $\tilde{\Psi}$, as a function of dimensionless shock position $x_s$. $\tilde{\Psi}$ is an integral measure for the balance of pressure, gravity, and ram pressure.  \citet{MURPHY2017} showed that $\tilde{\Psi}(x_s) = 0$ is a condition for a stalled shock solution, $\tilde{\Psi} < 0$ corresponds to $\varv_s < 0$, and $\tilde{\Psi} > 0$ corresponds to $\varv_s > 0$.  When $\tilde{\Psi}_{\rm min} > 0$ there are only $\varv_s > 0$ solutions, which corresponds to a critical condition for explosion.  Each line shows a different value of the dimensionless internal energy, $\tilde{\varepsilon}_{\rm NS}$, and a fixed value of $\tilde{\kappa}=0.52$. As $\tilde{\varepsilon}_{\rm NS}$ increases, the $\tilde{\Psi}$ curve shifts up and at the critical value  $\tilde{\varepsilon}_ c$, the minimum of the curve becomes zero. Below $\tilde{\varepsilon}_c$ a stalled shock solution exists, and above $\tilde{\varepsilon}_c$ only $\varv_s > 0$ solutions exist. The goal is to derive an analytic condition for the critical heating parameter for explosion, $\tilde{\varepsilon}_c$.}
\label{PsiVsXs}
\end{figure}

Since neutrinos are important in transporting energy and lepton number, two more physical dimensionless parameters are the dimensionless neutrino luminosity, $\tilde{L}_{\nu}$, and the dimensionless neutrino opacity, $\tilde{\kappa}$:  
\begin{align}
    \tilde{L}_\nu&=\frac{L_{\nu} R_{\rm NS}}{\mid \dot{M} \mid G M_{\rm NS}}  \label{eq_tilde_L}\, ,\\
    \tilde{\kappa}&=\frac{\kappa \mid \dot{M} \mid}{\sqrt{ G M_{\rm NS} R_{\rm NS}}} \, .
\label{eq_tilde_kappa}
\end{align}
The neutrino opacity, $\kappa$, is mostly constant and is set by microphysics. However, the mass accretion rate, $\dot{M}$, the mass of the neutron star, $M_{\rm NS}$, and the radius of the neutron star, $R_{\rm NS}$ are not constant, which makes $\tilde{\kappa}$ an important dimensionless parameter of the problem. 
Later, in section~\ref{deriving analytic critical condition}, we show that $\tilde{L}_{\nu}$ and $\tilde{\kappa}$ are two important parameters in the explosion condition.  $\tilde{L}_\nu$ compares the neutrino luminosity with the accretion power.  At first glance, $\tilde{\kappa}$ is not as intuitive, however, we now show that it parameterizes the neutrino optical depth for matter that has accreted onto the NS in a free-fall time.

The opacity is $\kappa = \sigma/m_p$.  We also note that the denominator of eq.~(\ref{eq_tilde_kappa}) is $\sqrt{GM_{\rm NS} R_{\rm NS}} \approx R_{\rm NS}^2/t_{\rm ff}$. Hence,
\begin{equation}
\tilde{\kappa} \approx \frac{\dot{M} t_{\rm ff}}{m_p} \frac{\sigma}{R_{\rm NS}^2} \approx \frac{4 \pi \Delta N_b}{N_{\sigma}} \, .
\end{equation}
$\Delta N_b$ is roughly the number of baryons accreted via $\dot{M}$ onto the NS surface in a dynamical timescale, $t_{\rm ff}$.  $N_\sigma$ is roughly the total number of cross-sectional areas within the surface area of the Neutron star, $N_\sigma = 4 \pi R_{\rm NS}^2/\sigma$. The ratio of $\Delta N_b$ to $N_{\sigma}$ gives the fraction of the NS surface that is opaque to neutrinos.  In other words, $\tilde{\kappa}$ is one minus the surface porosity to neutrinos.  This is similar to but not exactly the same as the neutrino opacity. Later, we show that this parameter is related to the density at the surface of the NS, or neutrino sphere.

Equation~(\ref{psi1}) represents an integral condition for the balance of forces.  As an integral condition, it self-consistently includes the boundary conditions.  One set of boundary conditions represents the shock jump conditions.  To facilitate analytic solutions we make the following assumptions for these boundary conditions.   For one, the pre-shock material is pressureless and in free fall.  Hence, $\tilde{p}_{+\epsilon} = 0$ and $\tilde{\varv}_{+\epsilon} = -\sqrt{2/x_s}$.  For core-collapse shocks, the compression ratio, $\beta = \rho_{-\epsilon}/\rho_{+\epsilon}$, is typically of order 7 to 9. 7 corresponds to the compression for a relativistic gas ($\gamma = 4/3$), and the higher compression ratio is due to photodissociation of heavy nuclei within the shock (\citet{FERNANDEZ2009}, \citet{MURPHY2017}). Given these approximations, the boundary conditions at the shock are:
\begin{eqnarray}
\tilde{\rho}_{+\epsilon}&=& \frac{x_s^{-3/2}}{4 \pi \sqrt{2}}\, , \\
\tilde{\varv}_{+\epsilon}&=& -\sqrt{\frac{2}{x_s}}\, , \\
\tilde{p}_{-\epsilon}&=& \tilde{\rho}_{+\epsilon}\tilde{\varv}_{+\epsilon}^2 \left ( 1 - \frac{1}{\beta}\right ) = \frac{x_s^{-5/2}}{2 \pi \sqrt{2}} \left ( 1 - \frac{1}{\beta}\right )\, .
\label{eq_p-epsilon}
\end{eqnarray}

Next, we approximate the three terms in $\Psi$, eq.~(\ref{psi1}).  The first term represents the the force due to ram pressure.  Integrating this term leads to a difference of the ram force at the shock minus the ram force at the NS surface. Simulations show that the boundary force term at the shock is typically $15$ times greater than the boundary force term at the surface of the neutron star. With these approximations, the ram pressure integral is
\begin{equation}
\begin{array}{rcl}
4 \pi \int^{x_s + \epsilon}_1 \left ( \frac{\partial}{\partial x} (\tilde{\rho} \tilde{\varv} x^2) \right ) dx & \approx & 4 \pi \tilde{\rho}_{+\epsilon} \tilde{\varv}_{+\epsilon}^2 x_s^2  \\
& \approx & \sqrt{2} x_s^{-1/2} \, . 
\end{array}
\label{first term in eq. Psi}
\end{equation}

The second term in eq.~(\ref{psi1}) represents a volume integral over the gradient of the pressure.  The $r^2$ factor presents a challenge in deriving an analytic expression for this term.  First, we divide the integral into two domains.  One domain is from the neutron star, $x = 1$, to just below the shock, $x_s - \epsilon$.  The second domain is from just below the shock, $x_s - \epsilon$, to just above the shock,  $x_s + \epsilon$.
\begin{equation}
4 \pi \int^{x_s + \epsilon}_1 \frac{\partial \tilde{p}}{\partial x} x^2 dx = 
4 \pi \left ( 
\int^{x_s - \epsilon}_1 \frac{\partial \tilde{p}}{\partial x} x^2 dx
+ \int^{x_s + \epsilon}_{x_s - \epsilon} \frac{\partial \tilde{p}}{\partial x} x^2 dx
\right )
\label{second term in eq. Psi_0}
\end{equation}
For the second domain, the $x^2$ factor is a constant, $x_s^2$, and the integral of the gradient results in the difference of the pressures below and above the shock.
\begin{equation}
\begin{array}{rcl}
4 \pi \int^{x_s + \epsilon}_{x_s - \epsilon} \frac{\partial \tilde{p}}{\partial x} x^2 dx &\approx& - 4 \pi x_s^2 \tilde{p}_{-\epsilon} \\
&\approx& -\sqrt{2} x_s^{-1/2} \left ( 
1 - \frac{1}{\beta}
\right ) \\
\end{array}
\end{equation}

The integral for the lower domain requires knowing the pressure profile.  Using the mean value theorem, we show that this term is generically proportional to the pressure at the NS and a dimensionless function of $x_s$. As long as the integrand is continuous and differentiable, then the mean value theorem states that there exists a point, $x_c$, in the interval such that:  

\begin{equation}
4 \pi \int^{x_s - \epsilon}_{1} \frac{\partial \tilde{p}}{\partial x} x^2 dx = 4 \pi  x_c^2  (\tilde{p}_{-\epsilon} - \tilde{p}_{\rm NS}) \, ,
\label{second term in eq. Psi_2}
\end{equation}
Further progress on this term requires analytic estimates for $x_c^2(x_s)$ and $\tilde{p}_{\rm NS}$.  As it turns out, one does not need to know the detailed functional form of $x_c^2$ to derive a critical condition.  Numerical solutions show that there is a value of $x_s$ for which there is a minimum in $x_c^2(x_s)$, and this is enough information to derive a critical condition; more on this later.  The analytic estimate for $\tilde{p}_{\rm NS}$ is also more involved, requiring analytic estimates for $\tilde{\varepsilon}_{\rm NS}$, heating, and cooling.  We will come back to addressing $x_c$ and $p_{\rm NS}$ in a moment.

Completing the discussion for $\tilde{\Psi}(x_s)$, we discuss the third term, the gravity term.  As in eq.~(\ref{second term in eq. Psi_0}), we divide the integration region into two regions.  The integral associated with the infinitesimally small region around the shock is zero because there are no derivatives in the integral and the integration region approaches zero.  Therefore, assuming $\kappa$ is constant, the gravity term is
\begin{equation}
    4 \pi \int^{x_s + \epsilon}_1 \tilde{\rho} dx = 4 \pi \int^{x_s - \epsilon}_1 \tilde{\rho} dx = 4 \pi \frac{\tau}{\tilde{\kappa}}\, ,
\label{third term in eq. Psi}
\end{equation}
where $\tau$ is the neutrino optical depth
\begin{equation}
\tau = \int^{R_s}_{R_{\rm NS}} \kappa \rho dr \, ,
\end{equation}
and $\tilde{\kappa}$ is defined by eq.~(\ref{eq_tilde_kappa}) and is a measure of the neutrino optical depth in matter that is accreted in a free-fall time.  The neutrino optical depth, $\tau$ represents a boundary condition and is $2/3$.

Combining the expressions for each term, eqs.~(\ref{first term in eq. Psi}),~(\ref{second term in eq. Psi_2}),~\&~(\ref{third term in eq. Psi}), the dimensionless expression for force balance is
\begin{equation}
    \tilde{\Psi}(x_s) = -\frac{\sqrt{2}x_s^{-1/2}}{\beta} - 4\pi  x_c^2 (\tilde{p}_{-\epsilon} - \tilde{p}_{\rm NS}) - 4\pi\frac{\tau}{\tilde{\kappa}} \, .
\label{Psi(xs)}
\end{equation}
Equation~(\ref{Psi(xs)}) shows that $\tilde{\Psi}$ is a function of dimensionless shock radius $x_s$.
Further progress requires an analytic estimate for $\tilde{p}_{\rm NS}$. First, we assume a gamma-law EoS to relate $\tilde{p}$ to $\tilde{\varepsilon}$:
\begin{equation}
    \tilde{p}_{\rm NS} = (\gamma - 1)\tilde{\rho}_{\rm NS}\tilde{\varepsilon_{\rm NS}} \, .
\end{equation}

\begin{figure*}
\includegraphics[scale=0.18]{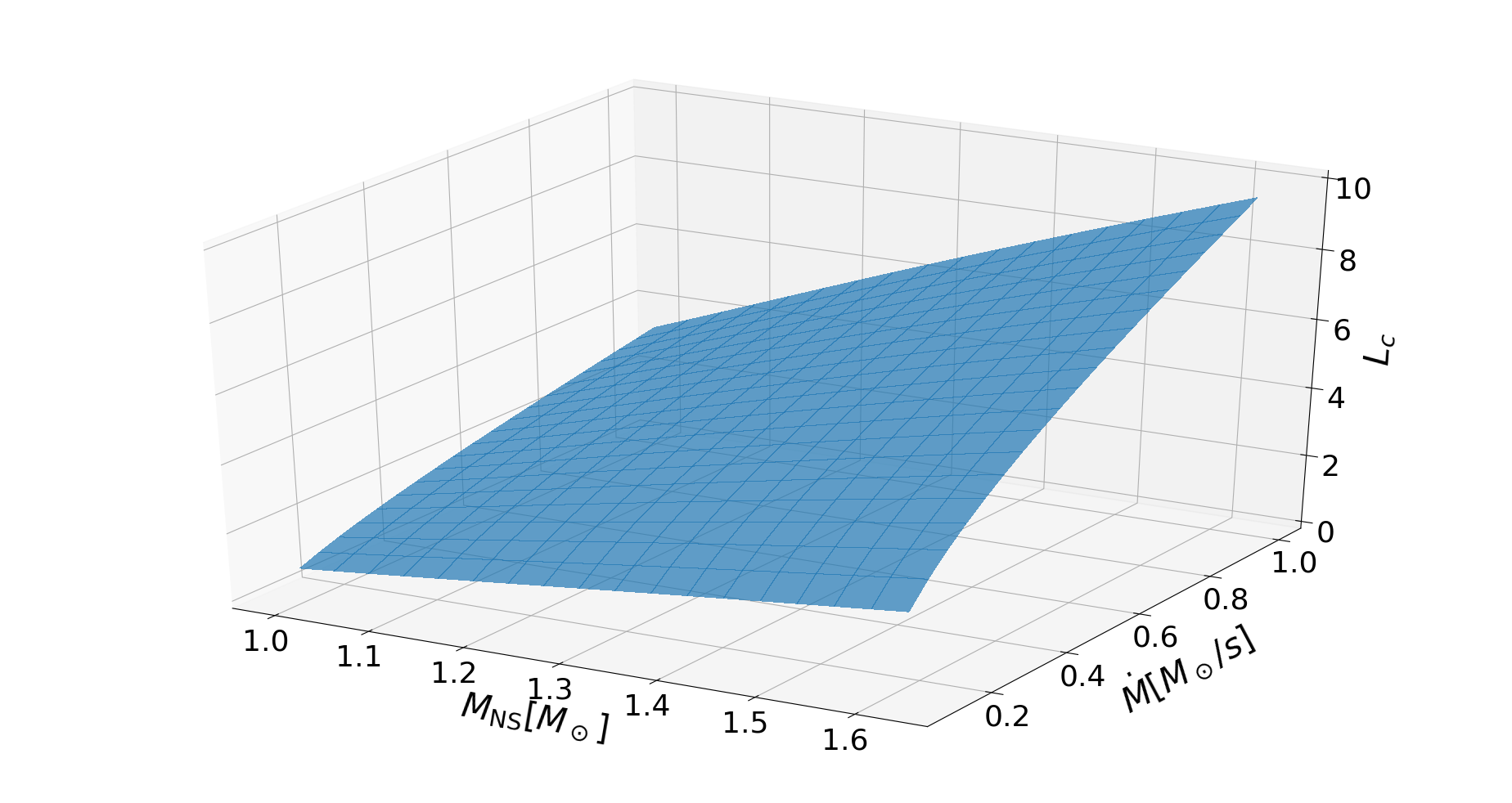}
\includegraphics[scale=0.18]{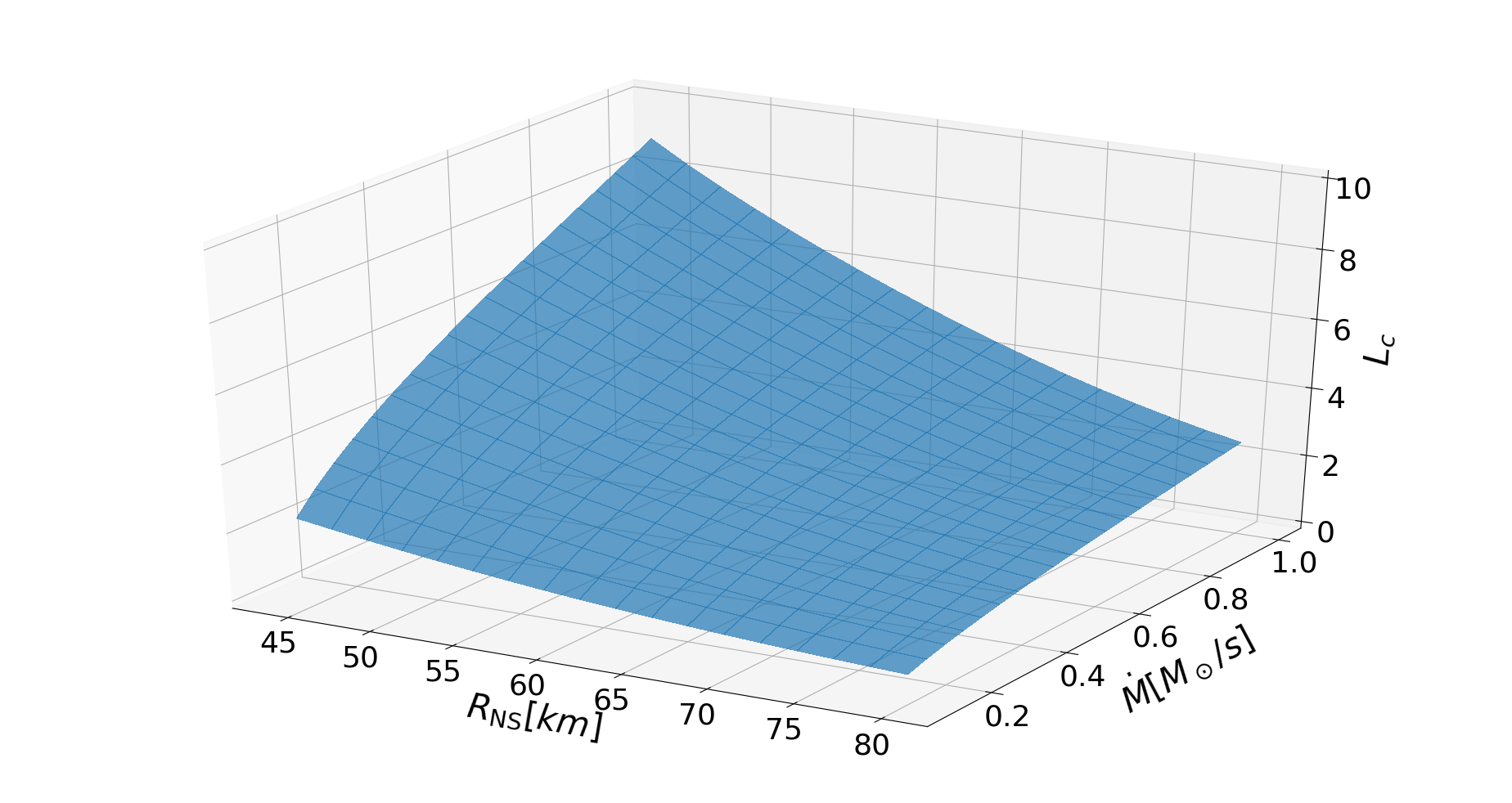}
\includegraphics[scale=0.18]{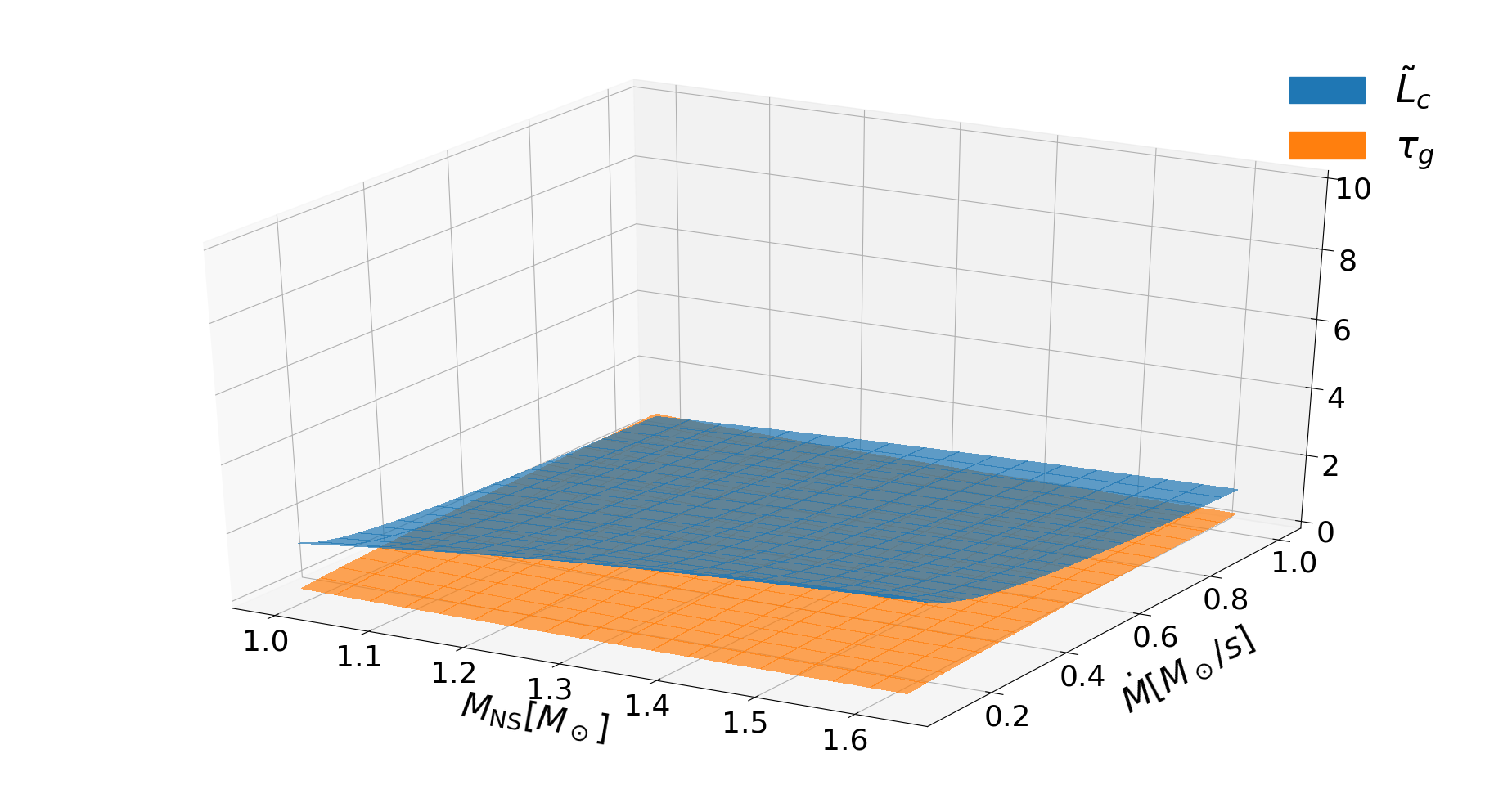}
\includegraphics[scale=0.18]{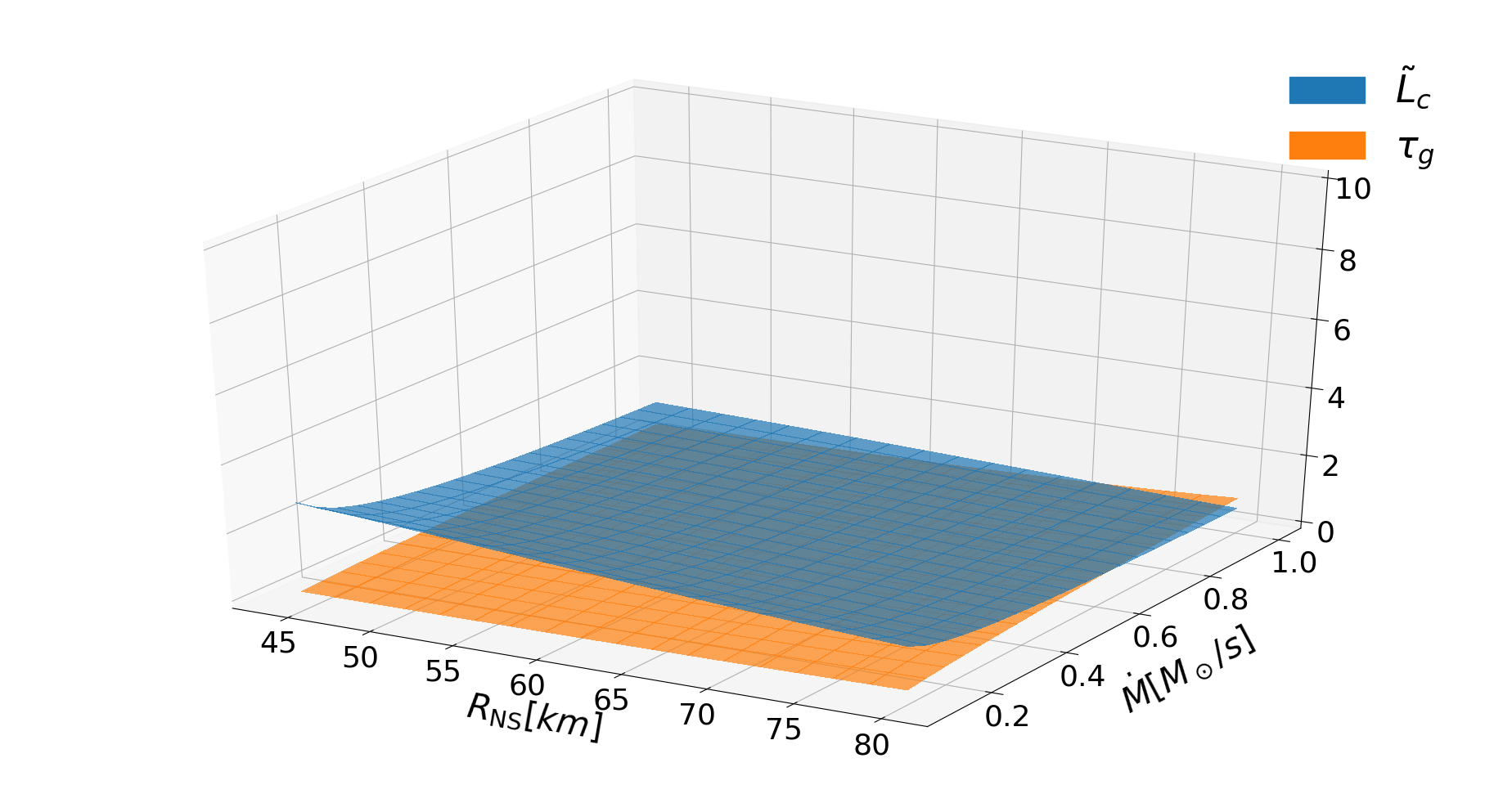}
\includegraphics[scale=0.18]{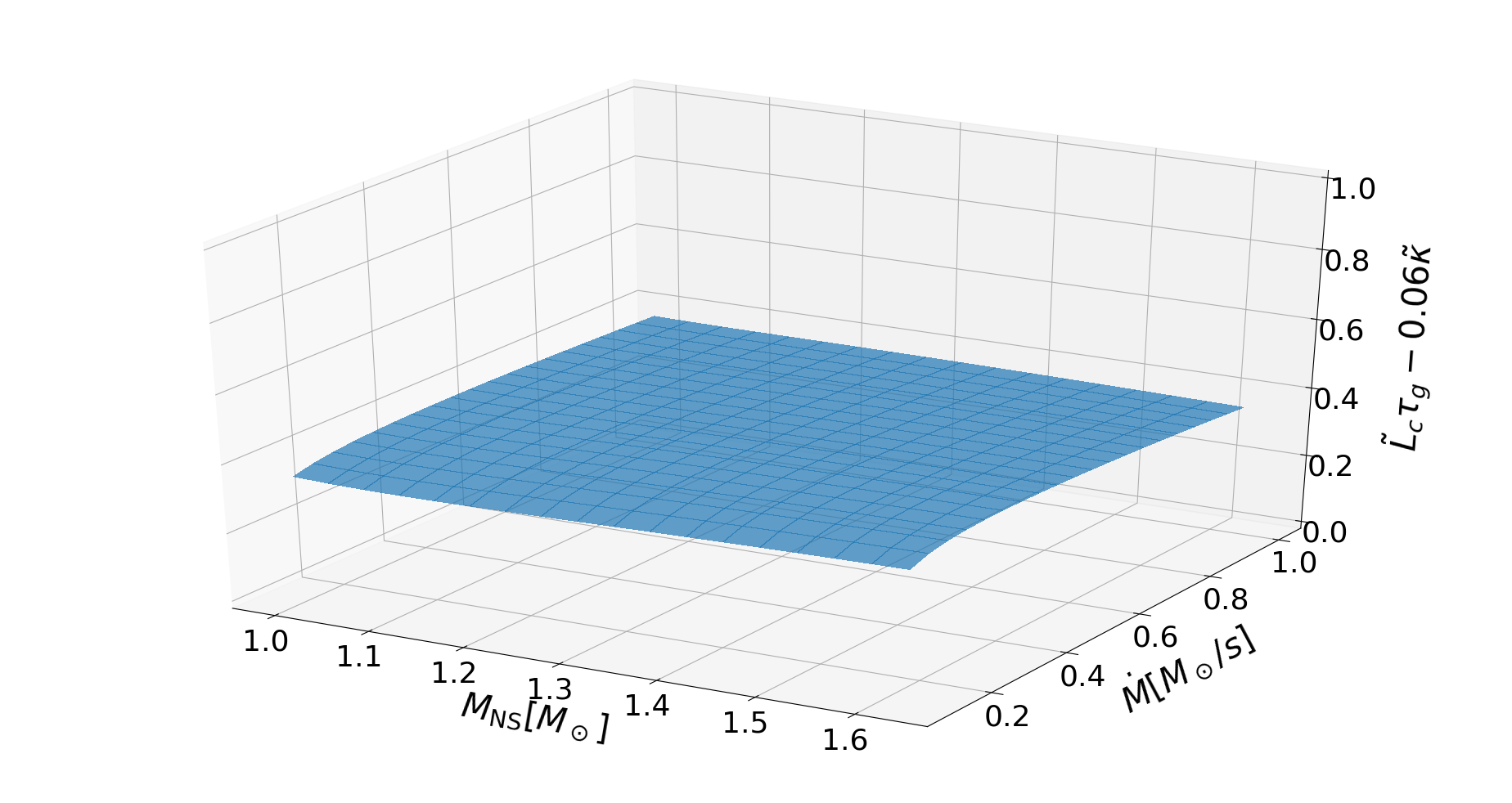}
\includegraphics[scale=0.18]{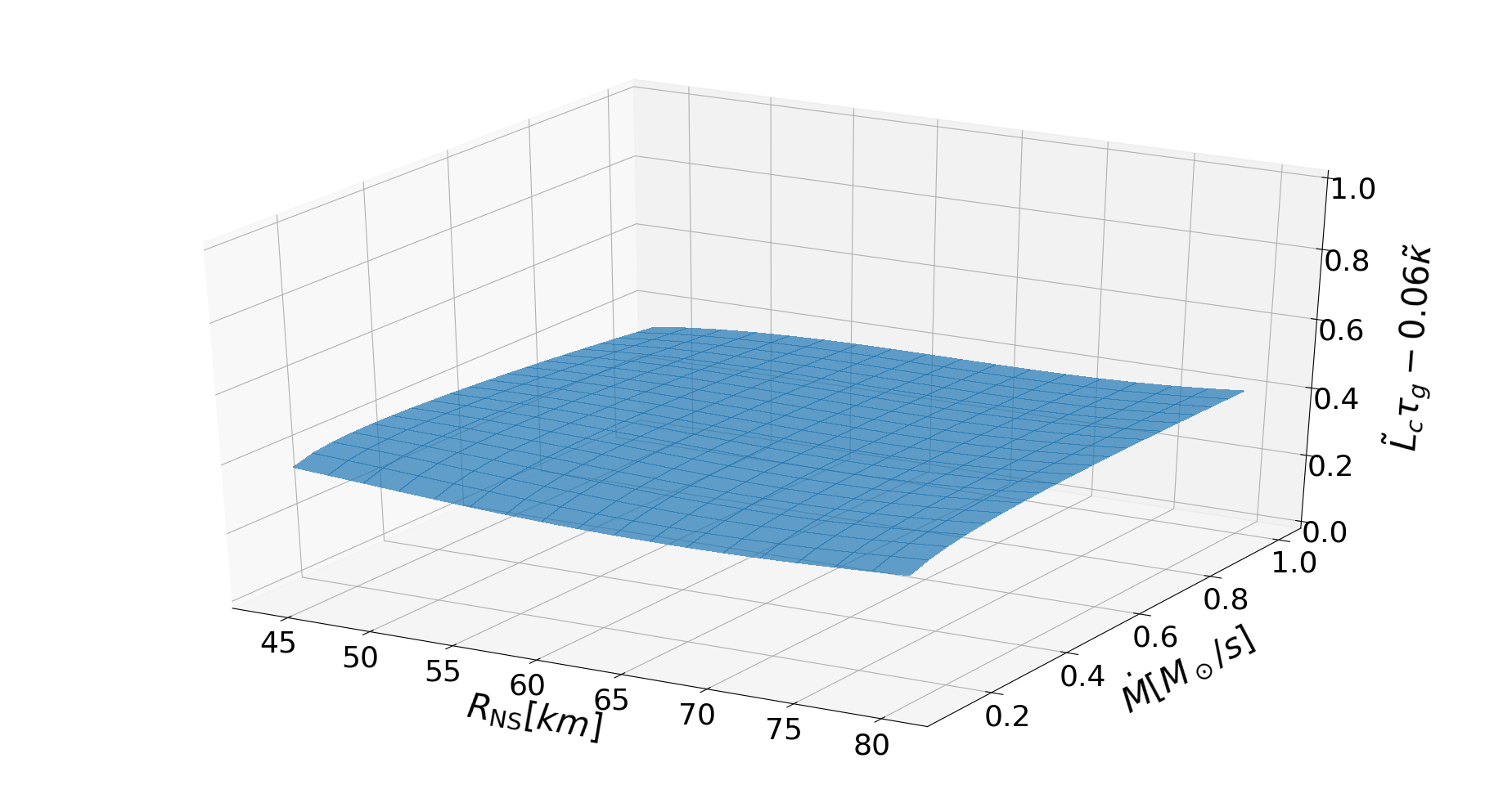}
\caption{A comparison of the analytic force explosion condition with numerical steady-state calculations. The top row shows the critical neutrino luminosity $L_c$ from the numerical calculations. The middle row shows the dimensionless critical neutrino luminosity $\tilde{L}_c$ and the neutrino optical depth of the gain region $\tau_g$. The bottom row shows the dimensionless measure of the critical neutrino heating per unit time in the gain region $\tilde{L}_c \tau_g$. The left column compares the critical conditions as a function of mass accretion rate $\dot{M}$ and neutron-star mass $M_{\rm NS}$,  and the right column compares these critical conditions as a function of $\dot{M}$ and the neutron-star radius $R_{\rm NS}$.   The critical neutrino luminosity (top row), $L_c$, varies by a factor of $\sim12$, the dimensionless neutrino critical luminosity varies by a factor of $5$, and the critical condition, $\tilde{L}_c \tau_g - 0.06\tilde{\kappa}$ varies by less than $\sim$20\%.  Since  $\tilde{L}_c \tau_g - 0.06\tilde{\kappa}$ is nearly constant, it provides the simplest and most intuitive critical condition for explosion.}
\label{surfs}
\end{figure*}

The density at the NS surface, $\tilde{\rho}_{\rm NS}$ is related to $\tau$ an important boundary condition of the steady-state solutions. \cite{BURROWS1993} showed that the structure between the NS and the shock is effectively a boundary value problem, where the NS sets the lower boundary and the shock jump conditions set the upper boundary.  \cite{BURROWS1993} proposed one boundary condition to be that $\tau=2/3$.  
To relate $\tilde{\rho}_{\rm NS}$ to $\tau$ we need a density profile.  In general, $\rho = \rho_{\rm NS} f(x)$.  Hence:
\begin{equation}
\tilde{\rho}_{\rm NS}=\frac{\tau}{\tilde{\kappa}F(x_s)} \, ,
\label{rho1}
\end{equation}
where $F(x_s)$ is the integral of $f(x_s)$ from the NS to the shock.  One may specify a functional form for $f(x_s)$.  However, we find it more informative to keep $f(x_s)$ as general a possible.  In fact, we will show that the concept of a critical condition does not depend upon the exact functional form of $f(x_s)$ but rather the general character of it.

\begin{figure*}
\includegraphics[scale=0.45]{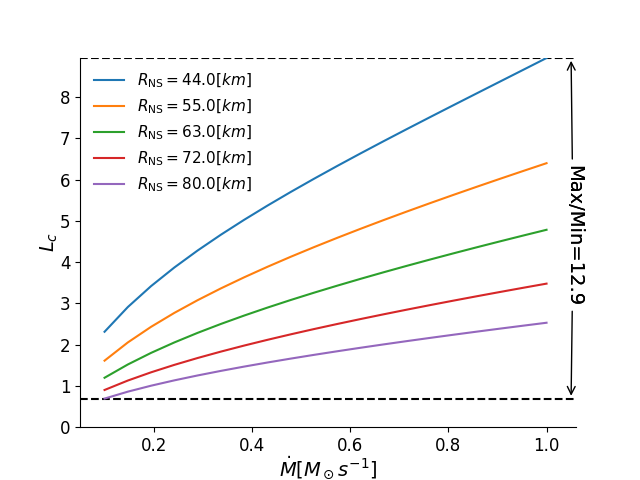}
\includegraphics[scale=0.45]{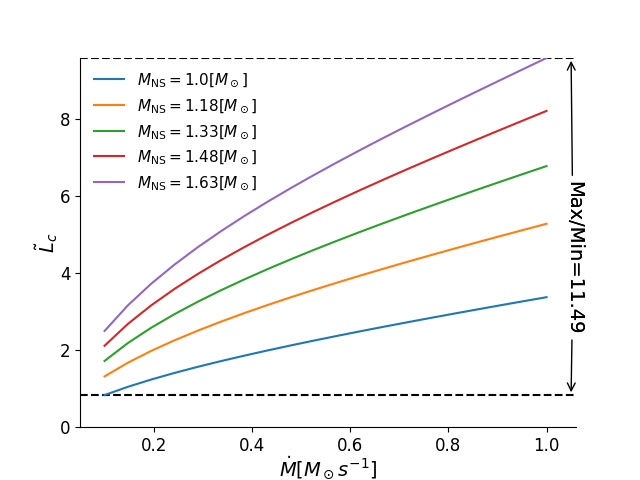}
\includegraphics[scale=0.45]{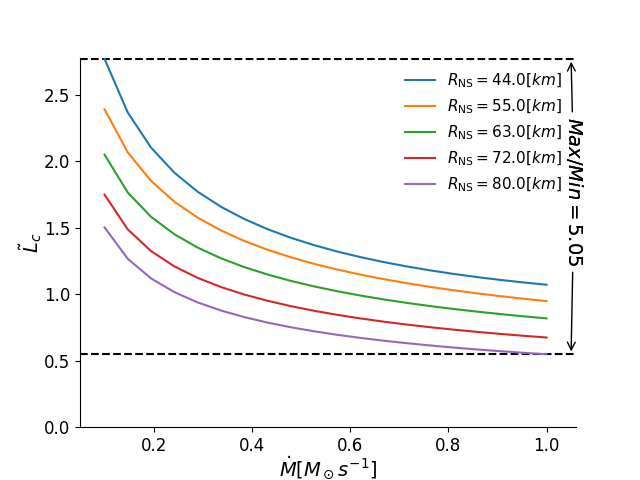}
\includegraphics[scale=0.45]{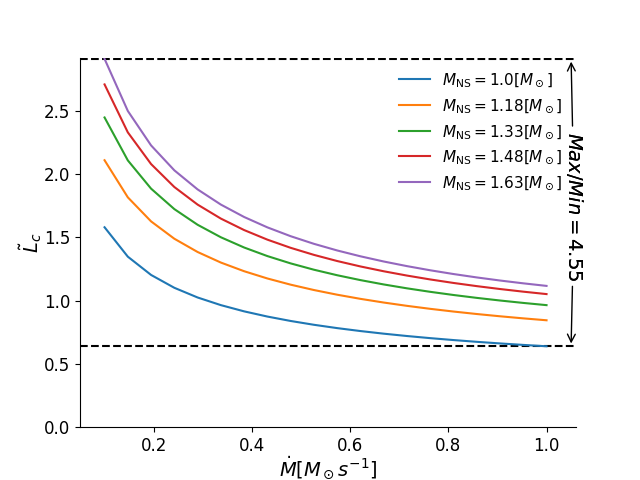}
\includegraphics[scale=0.45]{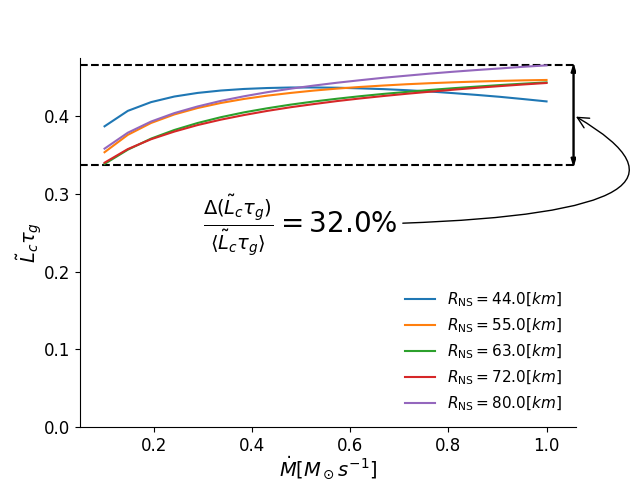}
\includegraphics[scale=0.45]{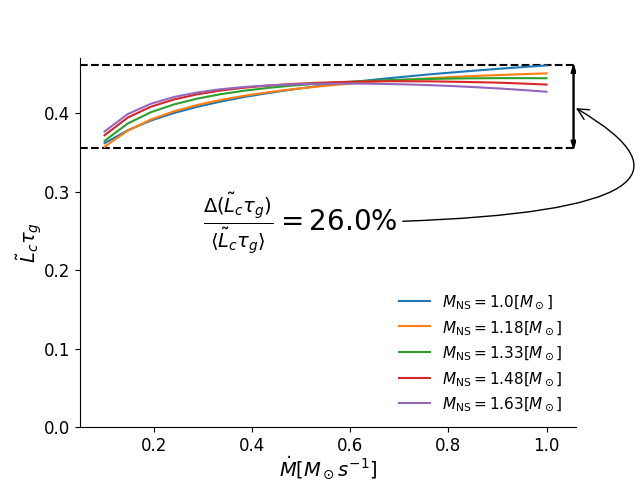}
\includegraphics[scale=0.45]{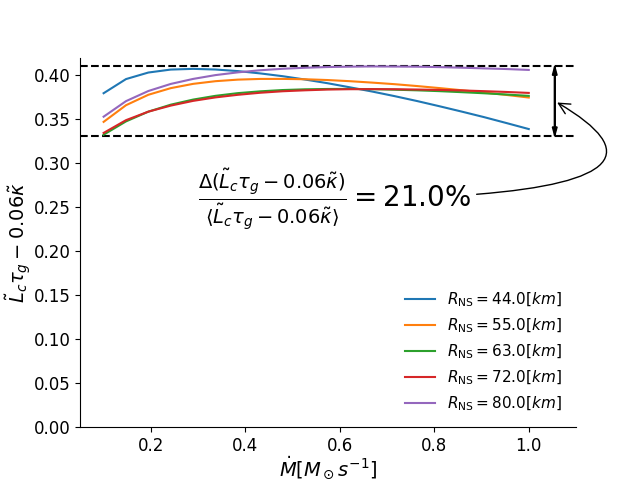}
\includegraphics[scale=0.45]{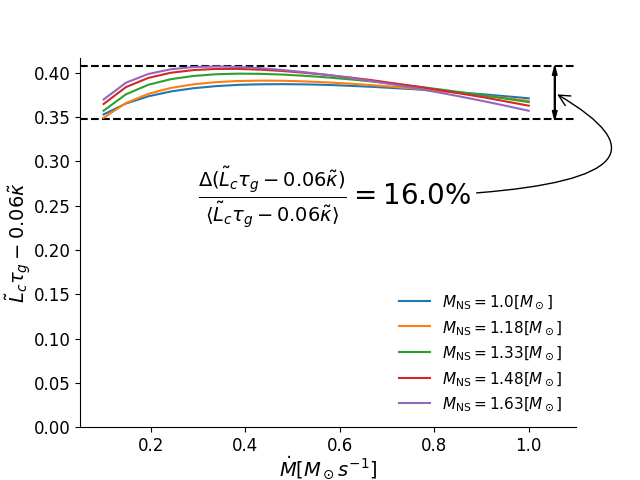}
\caption{Similar to Figure~\ref{surfs}, this figure shows a comparison of critical conditions.  The primary difference is that these panels show the critical conditions as a function of one parameter, $\dot{M}$.  In the left column, the neutron-star mass is fixed to $M_{\rm NS}=1.4 M_\odot$, and  each line represents a particular neutron-star radius, $R_{\rm NS}$.  In the right column, the neutron-star radius is fixed to $R_{\rm NS}=50 km$, and each line corresponds to a neutron star mass, $M_{\rm NS}$.  The third row shows just $\tilde{L}_c \tau_g$ while the bottom row shows the full force condition $\tilde{L}_c \tau_g - 0.06\tilde{\kappa}$.  Comparing these two rows shows that $\tilde{\kappa}$ accounts for some of the variation of the force explosion condition.  Again, the primary conclusion is that $\tilde{L}_c \tau_g - 0.06\tilde{\kappa}$ is the simplest and most accurate critical condition.}
\label{L_tL_heat_vs_Mdot}
\end{figure*}

To complete the analytic expression for $\tilde{p}_{\rm NS}$, we need an analytic approximation for the internal energy at the surface, $\tilde{\varepsilon}_{\rm NS}$.  The energy equation, eq.~(\ref{energy cont}), shows that the internal energy at the surface of the neutron star depends upon a steady-state structure set by gravity, advective flux, and neutrino heating and cooling. 
One may re-write eq.~(\ref{energy cont}) in terms of a partial differential equation for the temperature instead of internal energy.  This facilitates an approximate derivation that within the cooling region, neutrino cooling is balanced by neutrino heating and gravitational heating.  This in turn enables an analytic approximation for the internal energy.

The exact partial differential equation for temperature in spherical symmetry is 
\begin{equation}
    \bigg [\gamma \frac{\partial \varepsilon}{\partial \rho}\bigg\vert_{T}\frac{\partial \rho}{\partial r}+\gamma \frac{\partial \varepsilon}{\partial T}\bigg\vert_{\rho}\frac{\partial T}{\partial r}+\frac{\partial}{\partial r}\left(\frac{\varv^2}{2}\right)+\frac{\partial \Phi}{\partial r} \bigg] \frac{\dot{M}}{4\pi r^2}=\rho q_\nu-\rho q_c \, ,
\label{eq_exact_temp}
\end{equation}
where $\dot{M}={\rm const}$. 

\begin{figure*}
\includegraphics[scale=0.5]{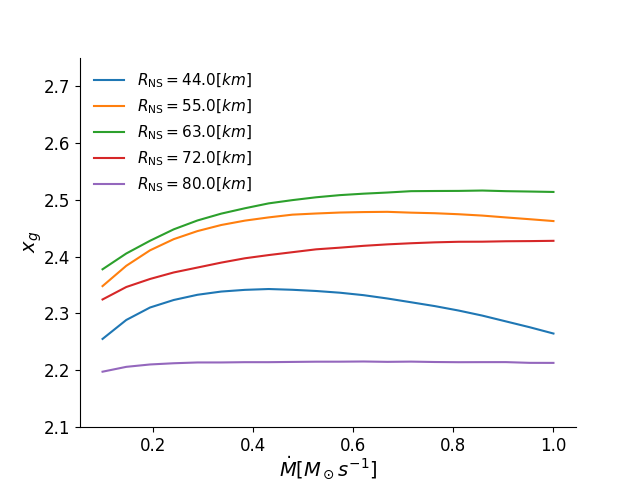}
\includegraphics[scale=0.5]{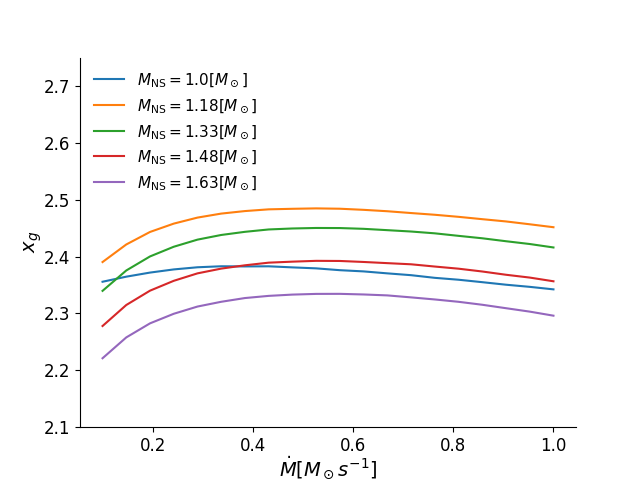}
\includegraphics[scale=0.5]{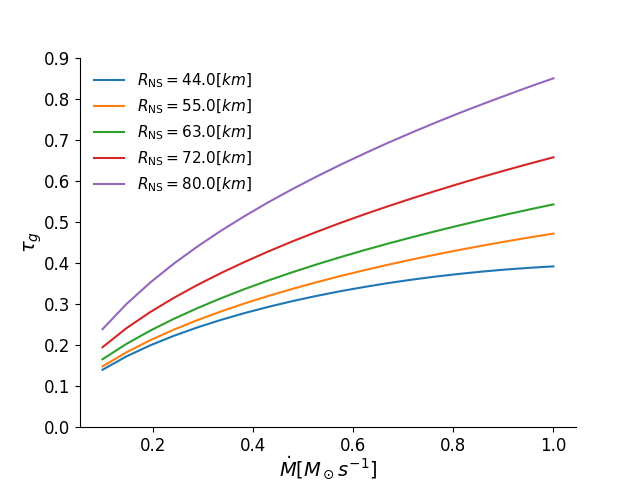}
\includegraphics[scale=0.5]{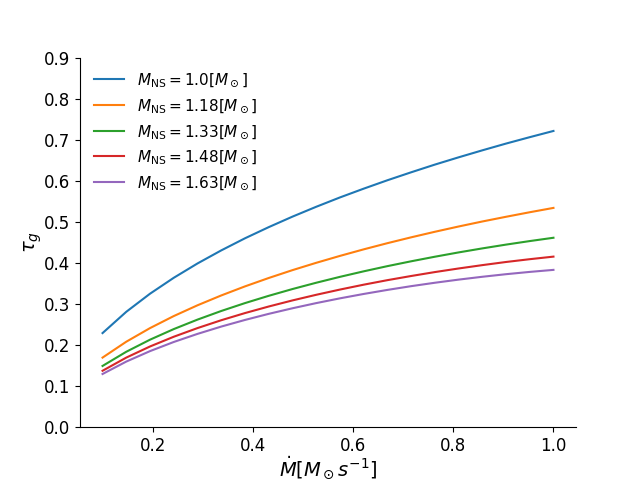}
\includegraphics[scale=0.5]{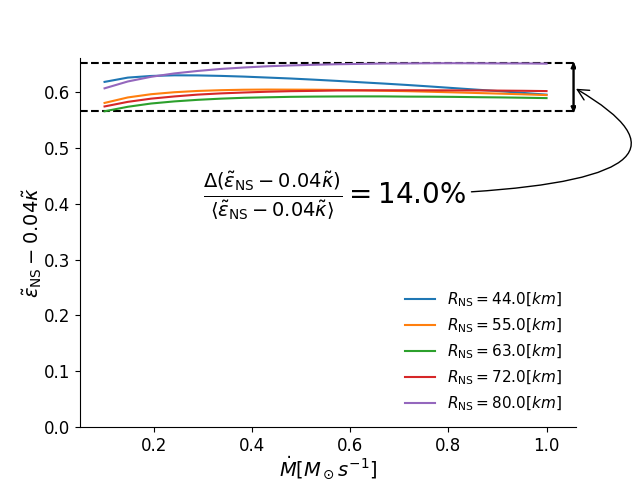}
\includegraphics[scale=0.5]{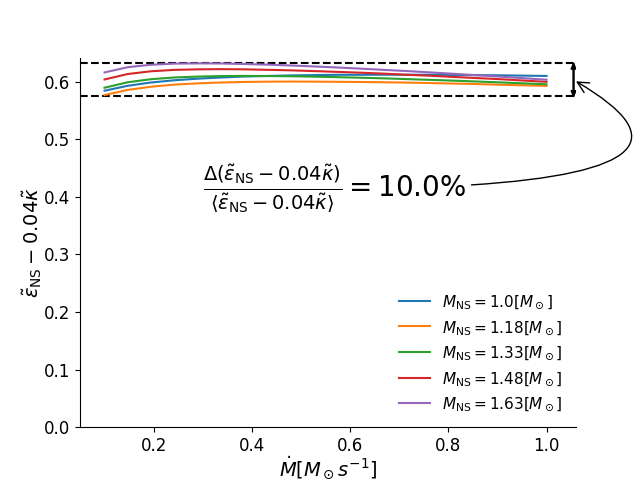}
\caption{A quantitative comparison for a related critical condition, $\tilde{\varepsilon}_{\rm NS}-0.04\tilde{\kappa} > \rm{const}$. The main difference here is that $\tilde{\varepsilon}_{\rm NS}$ includes the dimensionless gain radius $x_g$ in addition to $\tilde{L}_c\tau_g$.The top row shows the dimensionless gain radius $x_g$ as a function of mass accretion rate $\dot{M}$. The middle row shows the neutrino optical depth within the gain region, $\tau_g$. Finally, the bottom row shows the critical condition $\tilde{\varepsilon}_{\rm NS}-0.04\tilde{\kappa}$. In the left column, the mass of the neutron star is fixed, $M_{\rm NS}=1.4 M_\odot$, and each curve corresponds to a different value of $R_{\rm NS}$. In the right column, neutron star radius is fixed, $R_{\rm NS}=50$ km, and each curve represents a different value of $M_{\rm NS}$. Even though $\tilde{\varepsilon}_{\rm NS}-0.04\tilde{\kappa}$ has less variation than $\tilde{L}_c\tau_g-0.06\tilde{\kappa}$, it is  less intuitive. Since the only difference between the two is the relatively constant $1/x_g$ term, we highlight $\tilde{L}_c\tau_g-0.06\tilde{\kappa}$ as a critical condition. }
\label{xg_taug_H_vs_Mdot}
\end{figure*}

The material behind the shock is a combination of an ideal gas and radiation. Above the gain radius, radiation and relativistic particles tend to dominate the pressure and internal energy.  Below the gain radius, non-relativistic baryons and alpha particles tend to dominate the pressure and internal energy.  Therefore, within the cooling region we simplify eq.~(\ref{eq_exact_temp}), by assuming that the EoS is dominated by the ideal gas EoS, $\varepsilon = C T \propto T$ where $C$ is a proportionality constant. Furthermore, we assume that the velocity is small and ignore the kinetic term compared to the gravitational and thermodynamic energy densities. With these assumptions, the temperature profile within the cooling region roughly satisfies
\begin{equation}
\bigg(\gamma C \frac{\partial T}{\partial r}+\frac{\partial \Phi}{\partial r}\bigg)\frac{\dot{M}}{4\pi r^2}=\rho q_\nu -\rho q_c \, .
\label{eq_gradT}
\end{equation}
\begin{figure}
    \centering
    \includegraphics[scale=0.5]{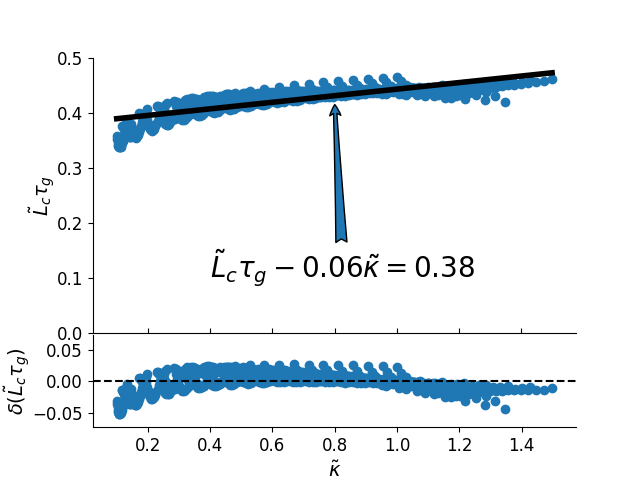}
    \caption{Correlation of  $\tilde{L}_c\tau_g$ and $\tilde{\kappa}$ from the numerical, steady-state critical condition. As is expected from the analytic derivation in section~\ref{deriving analytic critical condition}, the numerical results show that $\tilde{L}_c\tau_g$ is a linear function of $\tilde{\kappa}$. The solid line shows the linear fit to the numerical results. The bottom subplot shows the variation of $\tilde{L}_c\tau_g$ about the linear fit. The standard deviation of this variation is $0.014$. }
    \label{ProdVsKappa}
\end{figure}

To demonstrate the fidelity of these assumptions, Figure~\ref{TvsR} shows the temperature profile for a steady-state stalled solution.  As material settles onto the NS, it first passes through the gain region and then the cooling region.   Within the gain region, the temperature of the material increases as it advects downward.  This is partly due to gravitational heating (second term on the left-hand side of eq.~(\ref{eq_gradT}), which drives the temperature profile to $T \propto r^{-1}$, and partly due to neutrino heating.  The neutrino heating within the gain region makes the temperature profile slightly steeper than $1/r$.   Within the cooling region, the temperature gradient is roughly zero, which implies that neutrino cooling is roughly balanced by gravitational heating and neutrino heating.  This is a consequence of the steep dependence upon temperature, $q_c \propto T^6$.  There is a temperature where the neutrino cooling roughly balances the two heating sources.  If the temperature is below this temperature, then both types of heating will dominate.  Any temperatures above this temperature and the cooling term would dominate the heating  and quickly bring the temperature down to where cooling and heating balance again.  Therefore, the temperature gradient is roughly zero in the cooling region.  The rough balance between neutrino cooling and the two heating terms is:
\begin{equation}
q_c \approx q_\nu-\frac{G M_{\rm NS} \dot{M}}{4\pi \rho r^4} \, .
\label{cooling}
\end{equation}
Confirming this approximation, both numerical stalled solutions (Figure~\ref{TvsR}) and one-dimensional hydrodynamic solutions (Figure~\ref{fig_profiles} show a shallow temperature gradient within the cooling region.

To derive an analytic expression for the internal energy, we insert the approximation for cooling (eq.~(\ref{cooling})) into the steady-state equation for the internal energy, eq.~(\ref{energy cont}), and integrate from the NS to the shock (See details in the \ref{Appendix}).  The result is:
\begin{equation}
\tilde{\varepsilon}_{\rm NS}=\frac{1}{\gamma} \left( \tilde{B}(x_s)+\tilde{L}_\nu \tau_g+\frac{1}{x_g} \right) \, ,
\label{epsilon1} 
\end{equation}
where $\tilde{B}(x_s)$ is the Bernouli integral at the shock normalized by $G M_{\rm NS}/R_{\rm NS}$, and $\tau_g$ is the neutrino optical depth within the gain region:
\begin{equation}
\tau_g=R_{\rm NS} \int_{x_g}^{x_s}\kappa \rho dx \, .
\label{taug}
\end{equation}

While $\tilde{\varepsilon}_{\rm NS}$ seems to depend upon three parameters, it effectively depends upon only one.  $\tilde{B}$ is essentially zero, and the gain radius is a very weak function of the  physical parameters.  Hence, the dimensionless internal energy mostly depends upon one parameter, $\tilde{L}_\nu \tau_g$, the dimensionless neutrino power deposited within the gain region.

Finally, we have an analytic expression for integrated force balance:
\begin{equation}
\label{eq_tilde_psi}
\tilde{\Psi}(x_s) =- \frac{\sqrt{2}x_s^{-1/2}}{\beta} - 4 \pi x_c^2 (\tilde{p}_{-\epsilon} - (\gamma -1) \tilde{\rho}_{\rm NS} \tilde{\varepsilon}_{\rm NS}) - 4 \pi \frac{\tau}{\tilde{\kappa}} \, ,
\end{equation}
where $x_c^2$ and $\tilde{p}_{-\epsilon}$, eq.~(\ref{eq_p-epsilon}), are functions of $x_s$ only. $\tilde{\rho}_{\rm NS}$ is a function of $x_s$ and the ratio $\tau/\tilde{\kappa}$, and $\tilde{\varepsilon}_{}$ is effectively a parameter of the analytic equation.
 
Since one of the boundary conditions demands that $\tau = 2/3$, $\tilde{\Psi}$ is a function of two parameters: $\tilde{\varepsilon}_{\rm NS}$ and $\tilde{\kappa}$. In turn, the predominate free parameter within $\tilde{\varepsilon}_{\rm NS}$ is $\tilde{L}_{\nu} \tau_g$.  The upcoming derivation will show that the critical condition for explosion only depends upon these two parameters: $\tilde{L}_{\nu}\tau_g$ and $\tilde{\kappa}$. 

Note that eq.~(\ref{eq_tilde_psi}) is general and does not require particular assumption for density profile. \citet{MURPHY2017} used steady-state numerical solutions to show that $\tilde{\Psi}(x_s)$ has a minimum at some value of $x_s$. They also showed that the existence of this minimum is what leads to a critical condition.  When the minimum is below zero, i.e. $\tilde{\Psi}_{\rm min} < 0$, then there exists a shock radius, $x_s$ for which there is a stalled shock solution, i.e. $\tilde{\Psi}(x_s) = 0$.  If the minimum is above zero, then there is no stalled solution, but explosive solutions.  It is important to note that as long as there is a minimum in $\tilde{\Psi}$, then there is a critical condition.  In other words, we do not need to know the exact functional form for $\tilde{\Psi}(x_s)$, we just need to know that there is a minimum, and we need to know what sets the value of that minimum.  The dominant parameters in eq.~(\ref{eq_tilde_psi}) that set the scale of $\tilde{\Psi}$, and hence the minimum, are $\tilde{L}_{\nu}\tau_g$ and $\tilde{\kappa}$.  
Therefore, to find a critical condition for explosion, we must find $\tilde{L}_\nu \tau_g$ and $\tilde{\kappa}$ such that $\tilde{\Psi}_{\rm min} = 0$.


\subsection{Deriving Analytic Critical Condition}\label{deriving analytic critical condition}
In this section, we derive an analytic condition for $\min(\tilde{\Psi}(x_s)) = 0$, however such a derivation can lead to a cumbersome high order polynomial.  Fortunately, there is one more approximation that leads to a much simpler derivation, and as we show in later sections, this simpler derivation reproduces the explosion condition of the numerical ODE and hydrodynamic solutions.  The following presents this simpler derivation.

If one assumes that the minimum is at $x_s=x_{\rm min}$, then one notices that $\tilde{\Psi}(x_{\rm min})>0$ is equivalent to
\begin{equation}
     \tilde{\varepsilon}_{\rm NS}-a\tilde{\kappa} > b \, ,
\label{eq_crit_cond}
\end{equation}
where the analytic estimates for $a$ and $b$ are 
\begin{align}\label{eq_coef_a}
a&=\frac{F(x_{\rm min})}{\tau (\gamma-1)}\bigg( \tilde{p}_{-\epsilon}-\frac{\tilde{\varv}_{+\epsilon}}{4\pi  x_c^2 \beta}\bigg) \, , 
\end{align}
\begin{align}\label{eq_coef_b}
b&=\frac{F(x_{\rm min})}{ x_c^2  (\gamma-1)} \, , 
\end{align}
where $x_c^2$  is evaluated at $x_c^2(x_{\rm min})$, and
\begin{equation}
    F(x_{\rm min})=\frac{\tau}{\tilde{\rho}_{\rm NS}\tilde{\kappa}} \bigg|_{x_{\rm min}}  \, .
\end{equation}

Equation~(\ref{eq_crit_cond}) and eqs.~(\ref{eq_coef_a})~\&~(\ref{eq_coef_b}) provide a physical understanding of the explosion condition and the coefficients.  The first term in eq.~(\ref{eq_crit_cond}) presents the internal energy (or pressure) at the surface of the NS. The second term represents a difference in the thermal pressure and the ram pressure at the shock (see the first line in the definition for $a$ in eq.~(\ref{eq_coef_a})).  The coefficient $b$ on the right hand side represents a dimensionless scale for gravity.  In other words, the explosion condition in eq.~(\ref{eq_crit_cond}) is another way to state that an imbalance of integrated forces leads to explosion.


One can dive further into the the expression for $\tilde{\varepsilon}_{\rm NS}$ to derive both a more physical and practical critical condition  for diagnosing CCSN simulations.  First of all,  notice that  $\tilde{\varepsilon}_{\rm NS}$ is composed of several terms: $\tilde{B}$, $\tilde{L}_\nu \tau_g$, and $1/x_g$.  The last term, $1/x_g$ is a measure of gravity at the gain radius. The Bernoulli constant, $\tilde{B}$, and the inverse of the gain radius, $1/x_g$, are relatively constant.  Hence, a simpler critical condition to consider is
\begin{equation}
\tilde{L}_\nu \tau_g -a\tilde{\kappa} > b
\label{eq_LTau_const}
\end{equation}
Note that $b$ in this equation also represents gravity, but it has a slightly different meaning (and value) than in eq.~(\ref{eq_coef_b}).  In this explosion condition, $b$ represents the difference in gravity between the NS surface and the gain radius. 
This critical condition is a relatively simple and accurate explosion condition.  Not only is this condition simpler, but it is also more directly connected to the important physical parameters of the problem.  It is also a relatively easy parameter to calculate from CCSN simulations, making it a useful explosion diagnostic for CCSN simulations.

In summary, either of the explosion conditions, eqs.~(\ref{eq_crit_cond})~or~(\ref{eq_LTau_const}), represent a force explosion condition.
 The origin of the force explosion condition is eq.~(\ref{eq_tilde_psi}), which is another form of the momentum equation, eq.~(\ref{momentum}). Hence, the explosion condition demonstrates the balance of the main forces in the problem. These forces are the pressure gradient, gravitational, and the ram pressure forces. Since the explosion condition is an integral condition, the coefficients in the explosion condition depend upon the physical quantities measured at the shock and at the surface of the NS. The coefficient in front of $\tilde{\kappa}$, $a$, is relatively small because the quantities measured at the shock are normalized to the density at the surface of the NS. The constant $b$ comes from the gravity term in eq.~(\ref{eq_tilde_psi}), and is expected to be of order unity.

The existence of an explosion condition only requires that $\tilde{\Psi}$ has a minimum; it does not require an explicit functional form for the density profile.  However, to illustrate $\tilde{\Psi}$ graphically, we must specify the density profile.  In the following we assume that
 $\rho=\rho_{\rm NS} x^{-3}$.  

Figure ~\ref{PsiVsXs} shows the analytic integrated force-balance curves, $\tilde{\Psi}$, for different values of $\tilde{\varepsilon}_{\rm NS}$ and a fixed value of $\tilde{\kappa}=0.52$ (this value corresponds to typical values of the following physical parameters: $\dot{M}=0.4 M_{\odot}$, $M_{\rm NS}=1.4 M_{\odot}$, and $R_{\rm NS}=50$km). For low values of $\tilde{\varepsilon}_{\rm NS}$ (or luminosities), $\tilde{\Psi}(x_s)$ crosses $0$, hence a stalled shock solution exists; in other words, there is a shock radius for which $\tilde{\Psi}(x_s) = 0$.    Despite the approximations in this work, the  analytic curves reproduce the qualitative behaviour of the ODE solutions in \cite{MURPHY2017}. As $\tilde{\varepsilon}_{\rm NS}$ increases , the curve shifts up, and at a critical value, $\tilde{\varepsilon}_{\rm NS}=\tilde{\varepsilon}_c$ , $\tilde{\Psi}(x_s=x_{\rm min})=0$. Note that we use $\tilde{\varepsilon}_c$ as a shorthand of $\tilde{\varepsilon}_{\rm NS,c}$. For all values less than $\tilde{\varepsilon}_c$, stalled solutions are possible.  The critical value marks the highest value of $\tilde{\varepsilon}_{\rm NS}$ for which there is a stalled solution.  For larger values of $\tilde{\varepsilon}_{\rm NS}$ there are no stationary solutions for the ODEs.

Figure~\ref{PsiVsXs}, shows that the location of the minimum of $\tilde{\Psi}$ is roughly constant, $x_{\rm min} \sim 1.5$. Using this value, the analytic coefficients are $a \approx 0.05 $ and $b\approx 0.52$.  In section~\ref{ODE}, we use steady-state solutions to fit for more accurate coefficients.

In the following sections, we verify and validate this analytic critical condition with using numerical steady-state solutions in section~\ref{ODE} and with one-dimensional hydrodynamic simulations in section~\ref{CUFE}.

\section{Verifying Analytic Critical Condition with Numerical Steady-state Critical Conditions }\label{ODE}

In this section, we test the precision and accuracy of the analytic force explosion condition in section~\ref{deriving analytic critical condition} by comparing it  with the semi-analytic solutions of \cite{BURROWS1993}. 

\begin{figure*}
\includegraphics[width=\columnwidth]{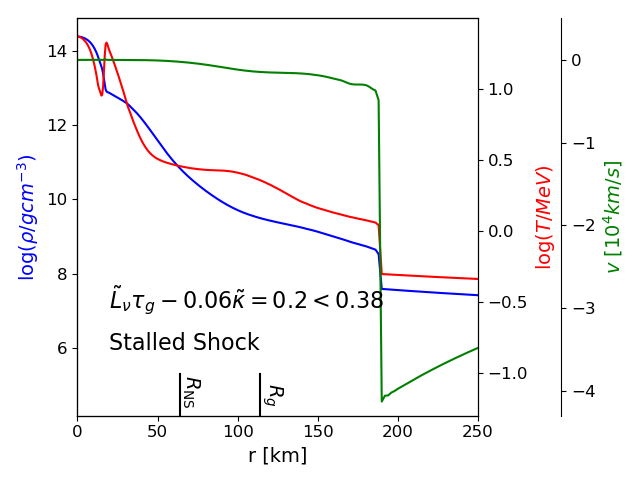}
\includegraphics[width=\columnwidth]{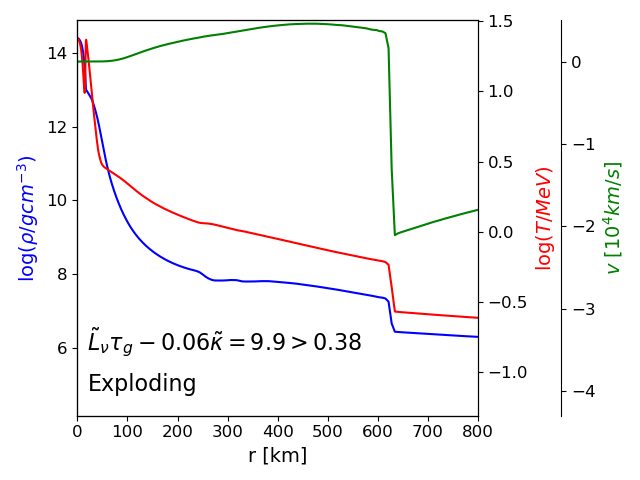}
\caption{Density, temperature, and velocity profiles of a one-dimensional hydrodynamic simulation  ($L_\nu=2.5$). The left panel highlights a stalled-shock phase (post bounce time is $350$ ms), and the right panel highlights a time after explosion begins (post bounce time is $820$ ms). For the stalled phase $\tilde{L}_\nu \tau_g - 0.06\tilde{\kappa} = 0.2$, which is less than the critical value for explosion, 0.38, and for the exploding phase $\tilde{L}_\nu \tau_g - 0.06\tilde{\kappa} = 9.9 > 0.38$.}
\label{fig_profiles}
\end{figure*}

First, we solve the steady-state hydrodynamic ODEs to find steady-state solutions,and we find the critical neutrino luminosity above which there are no steady-state solutions.  From hence forth, we will refer to this as the ODE-derived or semi-analytic critical neutrino luminosity.  For the most part, we follow the technique of  \citet{BURROWS1993}. They considered the CCSN problem as a boundary value problem with the lower boundary fixed at the surface of the proto-neutron star and upper boundary at the shock. They chose free parameters to be the neutrino luminosity $L_\nu$ and the mass accretion rate $\dot{M}$; the neutrino optical depth is fixed to be $2/3$. In addition to mass accretion rate, $\dot{M}$, we also vary the mass of the neutron star $M_{\rm NS}$, and the size of the neutron star $R_{\rm NS}$; To find the critical luminosity as a function of these parameters,  we vary the neutrino luminosity $L_\nu$ until we find the maximum neutrino luminosity for which there is a steady-state solution.  \cite{BURROWS1993} assumed that the solutions above are explosive and \cite{MURPHY2017} showed that indeed the only solution above this critical curve have a positive shock velocity solution. 

Figure~\ref{surfs} illustrates that the analytic force explosion condition of section~\ref{deriving analytic critical condition} is a natural explosion condition.  The two panels in Figure~\ref{surfs} show two projections of the critical neutrino-luminosity manifold. The left panel presents the critical neutrino luminosity as a function of mass accretion rate and the mass of the neutron star, and the right panel shows the critical luminosity as a function of the  mass accretion rate and the radius of the neutron star.  Within the range shown, the critical neutrino luminosity varies by a factor of $12$.  On the other hand, the middle two panels show two important dimensionless parameters: $\tilde{L}_c$ and $\tau_g$.  Their variation over the same region is significantly less: $5$ for $\tilde{L}_c$ and $6$ for $\tau_g$.  While these vary less than the critical neutrino luminosity, the bottom two panels show a more natural critical condition, $\tilde{L}_c\tau_g -0.06\tilde{\kappa}= \rm constant$, which varies by less than $20\%$.

Figure~\ref{L_tL_heat_vs_Mdot} also illustrates that the dimensionless critical condition, eq.~(\ref{eq_LTau_const}), is a more natural critical condition for explosion.   Figure~\ref{L_tL_heat_vs_Mdot} has four rows of panels.  The top row shows the dimensionful critical luminosity as a function of $\dot{M}$,$R_{\rm NS}$ (top-left panel) and $\dot{M}$,$M_{\rm NS}$ (top-right panel).  Similarly, the second row shows $\tilde{L}_c$ as a function of $\dot{M}$,$R_{\rm NS}$ ( second-left) and $\dot{M}$,$M_{\rm NS}$ (second-right), the third row shows $\tilde{L}_c\tau_g$ , and the bottom row shows $\tilde{L}_c \tau_g - 0.06\tilde{\kappa}$.  Plotting these quantities as a function of one variable better facilitates quantitative comparisons.  For example, $L_c$ varies by $\sim12$, $\tilde{L}_c$ varies $\sim 5$, $\tilde{L}_c \tau_g$ varies by less than 32\%, and $\tilde{L}_c \tau_g - 0.06\tilde{\kappa}$ varies by less than $20\%$.  While $\tilde{L}_c \tau_g$ captures much of the variation of the force condition, including the $\tilde{\kappa}$ term is a more precise explosion condition.

Figure~\ref{xg_taug_H_vs_Mdot} highlights an alternative analytic critical condition of $\tilde{\varepsilon}_{\rm NS}$.  The top row shows the dimensionless gain radius $x_g$ as a function of $\dot{M}$,$R_{\rm NS}$ (top-left panel) and $\dot{M}$,$M_{\rm NS}$ (top-right panel).  The middle row shows $\tau_g$ as a function of $\dot{M}$,$R_{\rm NS}$ (middle-left) and $\dot{M}$,$M_{\rm NS}$ (middle-right), and the bottom row shows $\tilde{\varepsilon}_{\rm NS} - 0.05\tilde{\kappa}$. While $\tilde{\varepsilon}_{\rm NS} - 0.04\tilde{\kappa}$ is slightly more accurate, $\tilde{L}_c\tau_g - 0.06\tilde{\kappa}$ is more intuitive and is a more practical explosion condition

The analytic derivation of the force explosion condition predicts that $\tilde{L}_c \tau_g$ linearly depends upon $\tilde{\kappa}$. 
Figure~\ref{ProdVsKappa} shows that the numerical steady-state results confirm this linear dependence.  When numerically evaluating $L_c$ from the steady-state equations, we choose a grid of values for $M_{\rm NS}$, $\dot{M}$, and $R_{\rm NS}$.  From these numerical results, we calculate $\tilde{L}_c \tau_g$ and $\tilde{\kappa}$ (blue dots).  The solid black line represents a linear fit.  The coefficient in front of $\tilde{\kappa}$ is $a = 0.06$, and the y-intercept is $b = 0.38$.  These coefficients are consistent with those derived from analytic estimates.  The bottom panel of Figure~\ref{ProdVsKappa} shows the difference between the numerical results and the linear fit.  The standard deviation about the fit is $0.014$, which represents a $4\%$ variation of the force explosion condition.

\section{Verifying Analytic Critical Condition with 1D light bulb Simulations} \label{CUFE}

\begin{figure*}
    \centering
    \includegraphics[scale=0.5]{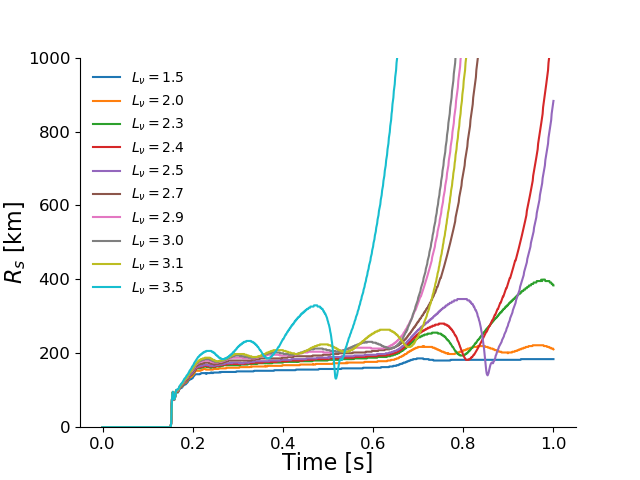}
    \includegraphics[scale=0.5]{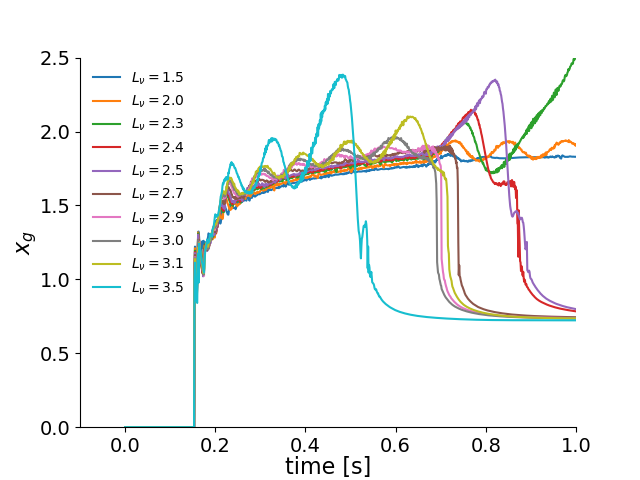}
    \includegraphics[scale=0.5]{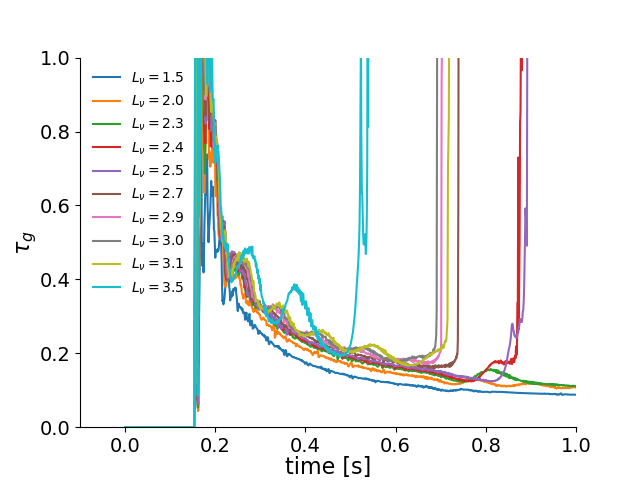}
    \caption{ Time evolution of shock radii (top panel), dimensionless gain radius, $x_g$, (middle panel), and the neutrino optical depth in the gain region, $\tau_g$, (bottom panel) for one-dimensional hydrodynamic simulations.  Each curve represents a different parameter for the neutrino luminosity. Simulations with $L_{\nu} > 2.3 \times 10^{52}$ erg/s explode within 1 second.  The oscillations in the shock radii and post shock structure are apparent in the gain radius and the optical depth.}
    \label{TaugVsTime}
\end{figure*}

In the previous section, we compared the analytic force explosion condition with the numerical steady-state condition; in this section, we compare the analytic explosion condition with hydrodynamic simulations that use the light bulb approximation for neutrino transport. For the one-dimensional simulations, we use  CUFE, which uses  Godunov's scheme to self-consistently solve hydrodynamic equations \citep{MURPHY2017}. In the following simulations, we simulate the collapse and explosion in spherical symmetry for a $12 M_\odot$ progenitor model \citep{WOOSLEY2007}. The EoS is a numerical table that includes effects of dense nucleons around and above nuclear densities, nuclear statistical equilibrium, electrons, positrons, and photons \citep{HEMPEL2012}. For the neutrino heating, cooling, and electron capture we use the light bulb approximation, which approximates neutrino transport with the
local descriptions of \cite{JANKA2001} and \cite{MURPHY2013}.  

Figure~\ref{fig_profiles} shows the density, temperature and velocity profiles from the simulation. The left panel corresponds to a time when the shock is stalled (post-bounce time of 350 ms). The value of the explosion parameter at this time is $\tilde{L}_\nu\tau_g-0.06\tilde{\kappa}=0.2$, which is below the explosion condition of $0.38$. The right panel shows a snap shot during explosion (post-bounce time of 820 ms).  The value of the explosion parameter at this time moment is $\tilde{L}_\nu\tau_g-0.06\tilde{\kappa}=9.9$, which is above the critical condition.

Figure~\ref{fig_profiles} also validates the assumption that neutrino cooling balances neutrino heating and gravitational heating in the cooling region (section~\ref{subsection: simple model}).  The temperature gradient (red line) is nearly zero in the cooling region ($R_{\rm NS} < r < R_{g}$).  $dT/dr \sim 0$ is an important approximation in deriving the balance in cooling and heating.  Again, this balance arises because the cooling term is a steep function of temperature ($\propto T^6$), and the cooling term brings the temperature down to the level of heating and no less.

Figure~\ref{TaugVsTime} shows the time evolution of the shock radii, dimensionless gain radius, $x_g=R_g/R_{\rm NS}$ (middle panel), and the optical depth within the gain region, $\tau_g$ (bottom panel) for different values of neutrino luminosity. The neutrino luminosities are in units of $10^{52}$ erg s$^{-1}$. The bounce time for these simulations is 150 ms. Simulations with $L_\nu \ge 2.3 \times 10^{52} \rm erg$ explode successfully. During the stalled phase, the dimensionless gain radius is relatively constant and is $x_g\approx 1.7$. The neutrino optical depth within the gain region, $\tau_g$, decreases over time, and for the simulations that do not explode, $\tau_g$ plateaus at $\sim 0.1$.  The reduction in $\tau_g$ is due to a nonlinear combination of a reduction $\dot{M}$, evolving radii, and neutrino luminosity.  We have yet to find a useful equation to relate $\tau_g$ to these other important parameters.  Since one can easily measure $\tau_g$ in the simulations, we leave $\tau_g$ as an important dimensionless parameter in the explosion condition.  In the future, it would be desirable to derive an analytic expression for $\tau_g$ in terms of the other important physical parameters.

\begin{figure}
    \centering
    \begin{subfigure}[t]{0.5\textwidth}
        \includegraphics[scale=0.5]{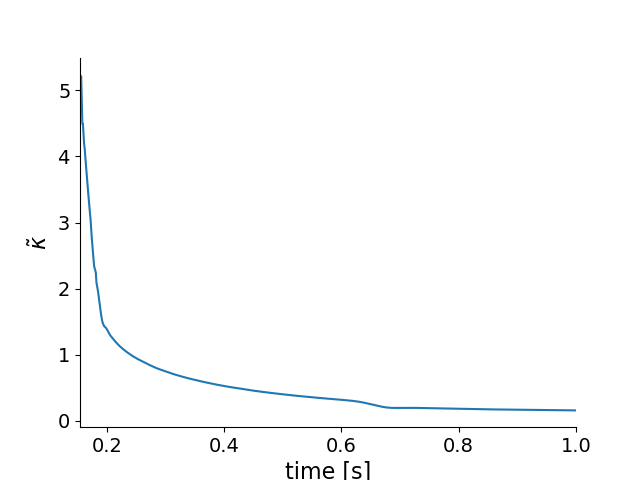}
    \end{subfigure}
    \begin{subfigure}[t]{0.5\textwidth}
        \includegraphics[scale=0.5]{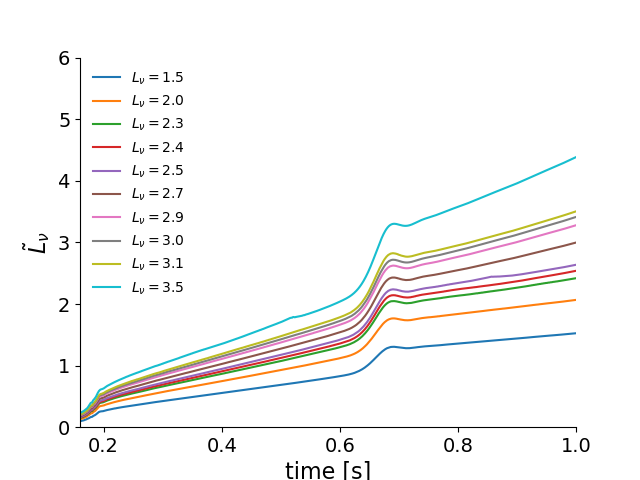}
    \end{subfigure}
    \caption{Time evolution of $\tilde{\kappa}$ (top panel) and $\tilde{L}_{\nu}$ (bottom panel).  These are two important dimensionless parameters of the critical condition. $\tilde{\kappa} = \frac{\kappa |\dot{M}|}{\sqrt{G M_{\rm NS} r_{\rm NS}}}$ parameterizes the neutrino optical depth of the accreted matter near the NS surface. $\tilde{L}_\nu =\frac{L_\nu}{|\dot{M}| \frac{G M_{\rm NS}}{r_{\rm NS}}}$ compares the neutrino luminosity to the accretion luminosity.  }
\label{TildeLVsTime}
\end{figure}

Figure~\ref{TildeLVsTime} shows the time evolution of $\tilde{L}_\nu$ and $\tilde{\kappa}$, the two important dimensionless parameters in the explosion condition.  $\tilde{L}_\nu$ compares the neutrino luminosity and accretion power, while $\tilde{\kappa}$ is a measure of the neutrino optical depth within the accreted material near the NS surface.
Both parameters require measuring $\dot{M}$, $M_{\rm NS}$, and $R_{\rm NS}$ as a function of time. Then, we  use eqs.~(\ref{eq_tilde_L})\&~(\ref{eq_tilde_kappa}) to calculate $\tilde{L}_\nu$ and $\tilde{\kappa}$. 

\begin{figure}
\includegraphics[width=\columnwidth]{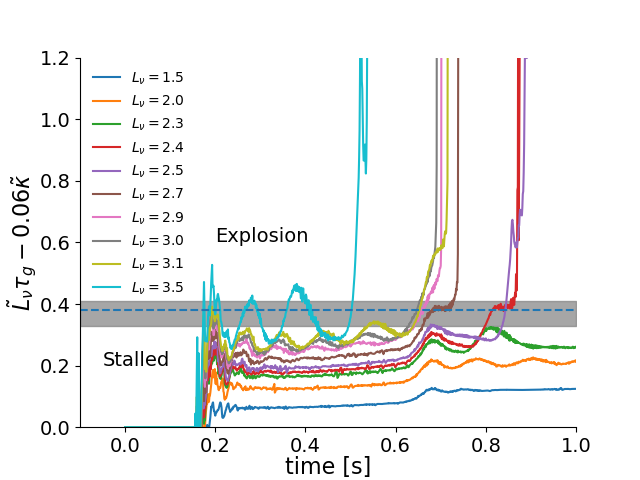}
\caption{ Time evolution of the force explosion condition, $\tilde{L}_\nu \tau_g - 0.06\tilde{\kappa}$. $\tilde{L}_\nu \tau_g$ compares the neutrino power deposited in the gain region with the accretion power, and $\tilde{\kappa}$ parameterizes the neutrino optical depth in the accreted matter at the neutron star surface. Each curve corresponds to one simulation with a specific neutrino luminosity.  The dashed line shows the best fit critical condition from the steady-state solutions (section~\ref{ODE} and Figure~\ref{surfs}). The gray band shows the variation in the critical condition based upon the steady-state solutions; see Figures~\ref{ProdVsKappa}~\&~\ref{L_tL_heat_vs_Mdot}.  This critical condition is relatively easy to calculate, accurately indicates when the simulations explode, and provides a useful indicator of how far away a simulation is from explosion.  The success of this one-dimensional test is the first step in verifying and validating that the force explosion condition is a valid and useful explosion diagnostic for more realistic, three-dimensional radiation hydrodynamic core-collapse simulations.}
\label{HeatVsTime}
\end{figure}

Finally, Figure~\ref{HeatVsTime} verifies and validates the force explosion condition with the one-dimensional light bulb simulations.  For each light bulb simulation we calculate the dimensionless expression $\tilde{L}_{\nu}\tau_g - 0.06\tilde{\kappa}$ as a function of time.  Our analytic derivation (section~\ref{deriving analytic critical condition}) and numerical fits (section~\ref{ODE}) suggest that $\tilde{L}_{\nu} \tau_g - 0.06 \tilde{\kappa} > 0.38$ is an explosion condition for spherically symmetric core-collapse supernovae.  The dashed blue line represents this explosion condition, and
 the gray band represents the bounds  of $\tilde{L}_\nu \tau_g-0.06 \tilde{\kappa}$ from Figure~\ref{L_tL_heat_vs_Mdot}. Our derivation states that below the gray band, stalled solutions exist, and above the gray band,  explosive solutions exist.  Indeed, the one-dimensional simulations tend to explode at or near the gray band.

\section{Discussion}\label{Discussion}

In this section, we derive the timescale, the ante-sonic, and the critical neutrino-luminosity conditions from the force explosion condition.

The timescale condition posits that explosion occurs if the advection time through the gain region is longer than the time it takes to heat that same material.  It is important to note that the timescale condition is not derived.  Rather, it is an heuristic condition.  The timescale condition is equivalent to stating that $\tilde{L}_{\nu} \tau_g > {\rm constant}$.  If one defines the advection timescale as $\tau_{\rm adv}  = M_g/\dot{M}$, and the heating timescale as $\tau_q=\frac{L_\nu \tau_g R_{\rm NS}}{G M_{\rm NS}M_g}$, where $M_g$ is the mass in the gain region, then this condition would imply that
\begin{equation}
\tilde{L}_{\nu} \tau_g = \frac{\tau_{\rm adv}}{\tau_q} \gtrsim {\rm constant} \, .
\end{equation}
Under these specific definitions for the timescales, then this explosion condition is close to the force condition.  However, it has two limitations.  For one, it is missing the $\tilde{\kappa}$ term.  While the coefficient of the $\tilde{\kappa}$ term is relatively small ($0.05$), the $\tilde{\kappa}$ term presents the effect of the upper boundary and the ram pressure on the force explosion condition.  Secondly, since the timescale condition is heuristic and not derived, there is no derivation for the constant in the timescale condition.


The ante-sonic critical condition \citep{PEJTA2012} for spherical explosions is $c_s^2/\varv_{\rm esc}^2 \simeq 0.19 $, where $c_s$ is the adiabatic sound speed and $\varv_{\rm esc}$ is the escape velocity at the shock.  We show that this condition is similar to stating that $\tilde{\varepsilon}_{\rm NS} > {\rm constant}$.
For a gamma-law EoS, the internal energy and sound speed are related by  $\varepsilon=\frac{c_s^2}{\gamma (\gamma-1)}$. 
If one ignores the $\tilde{\kappa}$ term, or ram pressure term in the force condition,  eq.~(\ref{eq_crit_cond}), 
\begin{equation}
    \tilde{\varepsilon}_{\rm NS} \gtrsim 0.6 \, .
\end{equation}
Rewriting eq.~(\ref{eq_crit_cond}) in terms of sound speed, $c_s$,  and escape velocity, $\varv_{\rm esc}$ gives
\begin{equation}
    \tilde{\varepsilon}_{\rm NS}=\frac{2 x_s^{-1}}{\gamma (\gamma-1)}\frac{c_s^2}{\varv^2_{\rm esc}} \gtrsim 0.6\, .
\end{equation}
The analytic solutions for $\tilde{\Psi}$ typically have minima around $x_{\rm min} = 1.5$ (Figure~\ref{PsiVsXs}). Hence, the approximate force explosion condition becomes 
\begin{equation}
    \frac{c_s^2}{\varv^2_{\rm esc}}\gtrsim 0.2 \, .
\label{eq_ante_sonic}
\end{equation}
This is remarkably similar to the explosion condition derived by \citet{PEJTA2012}.  There are three characteristics in the force explosion condition that are not in the ante-sonic condition.  For one, ante-sonic condition does not include the term associated with the ram pressure at the upper boundary.  Second,  \citet{PEJTA2012} derived the ante-sonic condition using an isothermal assumption.  Then using this as inspiration they posited that one would exist for adiabatic flows and fit for a possible critical value. 

Equation~(\ref{eq_ante_sonic}) shows that one can start with the FEC and (with some assumptions) derive the ante-sonic condition.  In Appendix B, \citet{PEJTA2012} reverse the derivation, and attempt to derive a critical luminosity condition based upon the ante-sonic condition.  They derive the following expression 
\begin{equation} 
L_{\rm \nu,core}^{\rm crit}\approx \eta \frac{GM\dot{M}}{r_\nu}\frac{1}{2\pi r_\nu \rho_\nu a} \big[1+\frac{\eta}{4}\big(1+\frac{1}{\eta}\frac{\varv_{\rm esc}\dot{M}}{\pi r_\nu \rho_\nu GM}\big)^2\big] \, ,
\label{eq:pejcha2012_1}
\end{equation}
where $\eta$ is the cooling efficiency.  It is more illuminating to re-write this expression in terms of the dimensionless parameters $\tilde{L}_\nu$ and $\tilde{\kappa}$:  
\begin{equation}
    \frac{\tilde{L}_\nu\tau}{2 \eta} - \frac{\eta}{4} \left ( 1 + \frac{\tilde{\kappa}}{\eta \tau \pi}\right )^2 \approx 1 \, .
    \label{eq:pejcha2012_2}
\end{equation}
While both the FEC derivation (eq.~(\ref{eq_LTau_const})) and the \citet{PEJTA2012} condition depend upon only two free parameters, there are several significant differences.  For one, rather than $\tau_g$, the total neutrino optical depth appears next to $\tilde{L}_\nu$.  
Second, eq.~(\ref{eq:pejcha2012_2}) has a quadratic dependence on $\tilde{\kappa}$, while the FEC has a linear dependence.   Granted while $\tau$ is a constant, $\tau_g$ might have some hidden dependence on $\tilde{\kappa}$.  However, comparing eq.~(\ref{eq:pejcha2012_2}) with the steady-state numerical solutions of \citet{BURROWS1993} shows that it does not reproduce the critical neutrino luminosity condition.  Similar to the analysis in Figure~\ref{ProdVsKappa}, we evaluate $\tilde{L}_\nu$ vs. $\tilde{\kappa}$.  Inspired by the quadratic form of eq.~(\ref{eq:pejcha2012_2}), we perform a quadratic fit:

\begin{equation}
    \tilde{L}_\nu \tau- 0.8 \tilde{\kappa}^2+1.72 \tilde{\kappa}=1.52
\label{eq:pejcha2012_fitting}
\end{equation}

One significant difference is that evaluating $\tilde{L}_{\nu} \tau$ as a function of ~$\tilde{\kappa}$ leads to significantly larger residuals compared to evaluating $\tilde{L}_{\nu} \tau_g$ vs.~$\tilde{\kappa}$.  For example, the residuals norm of eq.~(\ref{eq:pejcha2012_fitting}) is 0.22.  Where as the residuals norm of the FEC (linear fit for $\tilde{L}_{\nu} \tau_g$) is only 0.03.   Here we define the residuals norm as 
\begin{equation}
||e|| = \frac{\sqrt{\sum [y_i - f(\tilde{\kappa}_i)]^2/(N - k)}}{\overline{y_i}} \, , 
\end{equation}
where $y_i$ is either $\tilde{L}_{\nu} \tau$ or $\tilde{L}_{\nu} \tau_g$, $f(\tilde{\kappa})$ is either the linear or quadratic fit, $N$ is the total number of numerical evaluations, and $k$ is the number of fitting parameters.
In summary, the FEC condition has a factor of 10 smaller normalized residuals.

Another significant difference is in the estimates of the coefficients in eqs.~(\ref{eq:pejcha2012_2}) ~\&~(\ref{eq:pejcha2012_fitting}). If one assumes that $\eta \sim 0.5$, then eq.~(\ref{eq:pejcha2012_2}) gives $\tilde{L}_\nu \tau-0.11 \tilde{\kappa}^2-0.24 \tilde{\kappa}\approx 1.125$.  The estimated coefficients are a factor of 8 smaller, and the linear term has the opposite sign.  
As a final remark in this comparison, \citet{PEJTA2012} note that there is an inconsistency with an approximation in the derivation of Appendix B of \citet{PEJTA2012}.  It is possible that this inconsistency is responsible for the significant differences between the eq.~(\ref{eq:pejcha2012_2}) and the semi-analytic solutions of \citet{BURROWS1993}.


Deriving the critical neutrino luminosity \citep{BURROWS1993} from the force explosion condition is somewhat trivial.  As \citet{MURPHY2017} show, the critical neutrino luminosity and the force explosion condition represent the same critical hypersurface in the physical parameter space.  The critical neutrino luminosity curve is one projection of the critical hypersurface where $M_{\rm NS}$ and $R_{\rm NS}$ are held constant.  The primary difference of the force explosion condition is that it explicitly acknowledges and explores the dependence upon these other physical parameters.  Furthermore, we show that the force explosion condition is best represented by two dimensionless parameters, $\tilde{L}_{\nu} \tau_g$ and $\tilde{\kappa}$, and these parameters naturally explain how the force condition depends upon all of the physical parameters.

\section{Conclusions}\label{Conclusion}
    The primary result of this manuscript is an analytic derivation of the force explosion condition for spherically symmetric CCSNe, $\tilde{L}_\nu \tau_g-a\tilde{\kappa}>b$. When this condition is satisfied, the integrated thermal pressure overwhelms gravity and the ram pressure, leading to explosive solutions.  The two dimensionless parameters are $\tilde{L}_{\nu} = L_\nu R_{\rm NS} / (G M_{\rm NS} \dot{M})$, which compares the neutrino luminosity with the accretion power, and $\tilde{\kappa} = \kappa \dot{M} / \sqrt{G M_{\rm NS} R_{\rm NS}}$, which is a measure of the neutrino opacity for accreted matter near the neutron star surface.  In the explosion condition, the coefficient $a$ represents the difference between the thermal and ram pressure at the shock; Coefficient $b$ represents gravity.  Section~\ref{deriving analytic critical condition} provides analytic estimates for $a$ and $b$, but to ensure accurate predictions, we use stationary ODE solutions to fit for  $a$ and $b$. The numerical fits give $a \approx 0.06$ and $b \approx 0.38$  which are remarkably close to the analytic estimates. 
    
    The force explosion condition is an accurate and easily calculable explosion diagnostic for
     one-dimensional light-bulb simulations. Figure~\ref{HeatVsTime} most clearly demonstrates the consistency between the analytic force explosion condition and the hydrodynamic simulations. Below the gray band (analytic condition) the shock is stalled. Above the force condition, the hydrodynamic solutions are explosive.  In addition to indicating explosion, the force explosion condition provides a useful explosion diagnostic in that it quantifies how close the simulations are to explosion.
    
    The force explosion condition has promise to be an accurate and useful explosion diagnostic for three-dimensional radiation-hydrodynamic simulations.  To move toward this goal, we recommend the following tests and improvements.  While we used approximate light-bulb hydrodynamic simulations, the explosion condition will need to be tested with one-dimensional simulations in which neutrino transport is treated with an actual transport algorithm.  We suspect that one may simply replace $L_{\nu} \tau_g$ with the net integrated neutrino heating within the gain region.  To compare with multi-dimensional simulations, we propose to incorporate the mean-field convection model of \citet{MABANTA2018} into the explosion condition.  Then one may test whether the force explosion condition is consistent with three-dimensional radiation-hydrodynamic simulations.

\section*{Acknowledgements}
We thank the anonymous referee for careful reading of our
manuscript and insightful comments.

\section*{Data Availability}
The data underlying this article are available in the article.
\medskip
\bibliography{References}

\bsp
\appendix

\section{Internal energy at the surface of the NS}
\label{Appendix}
To evaluate the internal energy at the surface of the NS, we integrate eq.~\ref{energy cont} from the NS to the shock:
\begin{equation}
\int_{R_{\rm NS}}^{R_s} \boldsymbol \nabla \cdot \big(\rho \boldsymbol{\varv}(\epsilon+\frac{p}{\rho}+\frac{\varv^2}{2}+\Phi) \big) dV=\int_{R_{\rm NS}}^{R_s} (\rho q_\nu - \rho q_c) dV \, .
\label{int_eq_energy_con}
\end{equation}
Using $\gamma$-law EoS, $\varepsilon+p/\rho=\gamma \varepsilon$. We assume that the kinetic energy is small, hence ignore $\varv^2/2$ term in the left side of eq.~(\ref{int_eq_energy_con}). Moreover, $4\pi \rho \varv r^2=\dot{M}$. Therefore, left side of eq.~(\ref{int_eq_energy_con}) is
\begin{align}
&\int_{R_{\rm NS}}^{R_s} \boldsymbol \nabla \cdot \big(\rho \boldsymbol{\varv}(\epsilon+\frac{p}{\rho}+\frac{\varv^2}{2}+\Phi) \big) dV =\dot{M} \int_{R_{\rm NS}}^{R_s}\frac{d}{dr}(\gamma \varepsilon + \Phi)dr \nonumber \\ 
&=\dot{M} \int_{1}^{x_s}\frac{d}{dx}(\gamma \varepsilon + \Phi)dx = \dot{M}[B(x_s)-\gamma \varepsilon_{\rm NS}-\Phi_{\rm NS}] \, ,
\end{align}
where $B(x_s)$ is a Bernouli integral at the shock.

The region below the shock is divided into two regions: a neutrino-cooling dominated region below the gain radius, $R_g$ and a neutrino-heating dominated region above $R_g$.  These two regions exhibit qualitative differences that require different analytic approximations.  Hence, we divide the integration domain into a cooling region and a heating region.

To simplify right side of eq.~(\ref{int_eq_energy_con}), we divide the area of integration into two parts: 1) from $R_{\rm NS}$ to $R_g$ and 2) from $R_g$ to $R_s$. In the first region we approximate cooling using eq. ~(\ref{cooling}). In the second region neutrino heating is dominating, hence $\int_{R_g}^{R_s}\rho q_c dV$ is negligible. Therefore, the right side of eq.~(\ref{int_eq_energy_con}) is
\begin{align}
&\int_{R_{\rm NS}}^{R_s} (\rho q_\nu - \rho q_c) dV \approx \int_{R_{\rm NS}}^{R_s} \rho q_\nu dV - \int_{R_{\rm NS}}^{R_g} \rho q_c dV \nonumber\\ 
&=\int_{R_g}^{R_s} \rho q_\nu dV+\int_{R_{\rm NS}}^{R_g} \frac{GM_{\rm NS}\dot{M}}{4\pi r^4} dV \nonumber \\
&=L_\nu\tau_g-\frac{GM_{\rm NS}\dot{M}}{R_{\rm NS}}\bigg(\frac{1}{x_g}-1\bigg)
\end{align}
Finally, for $\varepsilon_{\rm NS}$ 
\begin{equation}
\varepsilon_{\rm NS}=\frac{1}{\gamma}\bigg(B(x_s)+\frac{L_\nu \tau_g}{|\dot{M}|}+\frac{GM_{\rm NS}}{R_{\rm NS}x_g} \bigg)    
\end{equation}
and hence
\begin{equation}
\tilde{\varepsilon}_{\rm NS}=\frac{1}{\gamma}\bigg(\tilde{B}(x_s)+\tilde{L}_\nu \tau_g+\frac{1}{x_g} \bigg)    
\end{equation}

\label{lastpage}
\end{document}